\documentstyle[psfig]{l-aa}
\begin{document}
\thesaurus{03 % A&A Section 3: Extragalactic Astronomy
(03.20.8;  % Techniques: spectroscopic,
11.04.1;  % Galaxies: distances and redshifts,
11.05.1;  % Galaxies: elliptical and lenticular, cD,
11.06.2)} % Galaxies: fundamental parameters.
\title{Fundamental Plane Distances to Early-type Field Galaxies in the South Equatorial Strip
\, I. The Spectroscopic Data}
\author{K. R. M\"uller\inst{1}\inst{2}
\and
G. Wegner\inst{1}
\and
W. Freudling\inst{2}}
\institute{Department of Physics and Astronomy, Dartmouth College, 6127 Wilder Laboratory,
Hanover NH 03755, USA
\and              
European Southern Observatory and Space Telescope -- European Coordinating Facility, 
Karl Schwarzschild Str. 2, D-85748 Garching bei M\"unchen, Germany}
\offprints{K. R. M\"uller}
\date{Received ....; accepted ....}
\maketitle
\markboth{M\"uller $et$ $al.$}{Spectroscopic Data}
%\maintitlerunninghead{Spectroscopic Data}
%\authorrunninghead{M\"uller $et$ $al.$}
\begin{abstract}
Radial velocities and central velocity dispersions are derived for 238 E/S0 galaxies from
medium-resolution spectroscopy.
New spectroscopic data have been obtained as part of a study of the Fundamental Plane
distances and peculiar motions of early-type galaxies in three selected directions of the South
Equatorial Strip, undertaken in order to investigate the reality of large-scale streaming motion;
results of this study have been reported in M\"uller $et$ $al.$ (1998).
The new APM South Equatorial Strip Catalog 
($-17^{\circ}.5 < \delta < +2^{\circ}.5$) was used to select the sample of field galaxies in three
directions: (1) 15h10 -- 16h10; (2) 20h30 -- 21h50; (3) 00h10 -- 01h30.
The spectra obtained have a median $S/N$ per $\mbox{\AA}$ of 23, an instrumental resolution
(FWHM) of $\sim$ 4 $\mbox{\AA}$, and the spectrograph resolution (dispersion) is $\sim$ 100
km~s$^{-1}$.
The Fourier cross-correlation method was used to derive the radial velocities and velocity
dispersions. The velocity dispersions have been corrected for the size of the aperture and for the
galaxy effective radius. 
Comparisons of the derived radial velocities with data from the literature show that our values
are accurate to 40~km~s$^{-1}$.
A comparison with results from J$\o$rgensen $et$ $al.$ (1995) shows that the derived central
velocity dispersion have an rms scatter of 0.036 in $\log \sigma$. There is no offset relative to
the velocity dispersions of Davies $et$ $al.$ (1987). 
\end{abstract}
% section 1
\section{Introduction}
The observed total radial velocity of a galaxy (found from the redshift) can be separated into two
distinct parts: a cosmological part, from the expansion of the universe, and a peculiar motion
associated with the galaxy proper motion. 
Measurements of galaxy peculiar velocities on large scales reveal the underlying mass density
fluctuations, since galaxies will stream towards an overdense region and away from an
underdense region.
 
To determine peculiar motions, a distance-indicator relation has to be used to find
redshift-independent distances to galaxies. The Tully-Fisher relation can be used for spiral
galaxies. Elliptical galaxies have been found to populate a nearly planar region in the
three-dimensional space defined by the central velocity dispersion, the effective
(half-luminosity) radius, and the effective surface brightness; this region is called the
Fundamental Plane (Lucey $et$ $al.$ 1991; J$\o$rgensen $et$ $al.$ 1993). The Fundamental
Plane (FP) method for distance determination is an improvement on the $D_n - \sigma$ relation.
It has a tighter correlation; therefore, a better precision in distances ($\sim$ 20\,\%) can be
achieved (J$\o$rgensen $et$ $al.$ 1996; Scodeggio $et$ $al.$ 1997).

Galaxy peculiar velocities are found from a comparison of the distances with the measured
redshifts. There is strong observational evidence for the existence of large-scale flows in the
local universe, induced by gravity (see Strauss \& Willick 1995). 
The dipole anisotropy of the cosmic microwave background (CMB) radiation provides a natural
velocity reference frame for the analysis of galaxy motions. The dipole anisotropy, determined
from COBE, implies that the Local Group (LG) moves with respect to the CMB rest frame at 627
$\pm$ 22 km~s$^{-1}$ towards $l = 276 \pm 3^{\circ}$, $b = +30 \pm 3^{\circ}$ (Kogut $et$
$al.$ 1993). If this has a kinematic origin then, sufficiently far away, galaxy peculiar velocities
should converge to the CMB frame. 
%Indeed, the observed LG motion relative to the restframe 
%defined by galaxies within 6000 km~s$^{-1}$ points towards 
%the Hydra Centaurus-Great Attractor region, which is within 
%$\sim 40^{\circ}$ of the CMB dipole direction.

Until now, the only studies which have reported measurements of the velocity field as far out
as 15~000 km~s$^{-1}$ are those of Lauer \& Postman (LP) (1994), using brightest cluster
galaxies as distance indicators, and Riess $et$ $al.$ (1995), using Type Ia supernovae. LP
checked the convergence of the LG dipole motion to the CMB dipole, with a surprising result:
a strong signature of a very large-scale bulk flow was seen, with an amplitude of 689 $\pm$ 178
km~s$^{-1}$ in the direction $l = 343^{\circ}$, $b = +52^{\circ}$. The LP study implies that
the local rest frame fails to converge to the CMB frame, even in regions with radii $\sim$
15~000 km~s$^{-1}$. A bulk flow with the statistical significance of this result rules out a
whole series of cosmological models at the $>95$\,\% confidence level (Feldman \& Watkins
1994; Strauss $et$ $al.$ 1995); the LP result is in disagreement with all viable models at present.

The LP sample extended to 15~000 km~s$^{-1}$, with an effective depth of $\sim 8000$
km~s$^{-1}$. Therefore, the logical next step was to compare the LP result with peculiar
velocities as found from applying the Tully-Fisher and FP methods to galaxies extending further
out than any previous peculiar velocity studies. From Tully-Fisher studies of field and cluster
spiral galaxies within 8000~km~s$^{-1}$, Giovanelli $et$ $al.$ (1996, 1998a, 1998b) concluded
that these galaxies do not show any evidence of such a bulk flow.
% also, ellipticals better because can go further out - add? **

In order to investigate the reality of large-scale streaming motion on scales of up to 150~Mpc,
we have studied the peculiar motions of 179 early-type galaxies in three directions of the South
Equatorial Strip, at distances out to $\sim$ 20~000~km~s$^{-1}$. 
We have obtained new and independent measurements of the peculiar velocity field of elliptical
field galaxies at a depth similar to that of LP, using a combination of photometric and
spectroscopic data.
For further details of the project, see M\"uller (1997); the results for peculiar motions are
analysed in M\"uller $et$ $al.$ (1998). 

In this paper we present the spectroscopic data used in our study of the large-scale motions. 
>From the spectra, galaxy redshifts were measured, and central velocity dispersions were 
obtained -- the accurate determination of these is essential for the FP to be applied as a distance
indicator. 
This paper is organised as follows. Sample selection and observations are described in Sect. 2,
the basic reduction of the spectra is covered in Sect. 3, and the radial velocities and central
velocity dispersions are derived in Sect. 4. The applied corrections are discussed in Sect. 5, and
the results are given in data tables in Sect. 6. In Sect. 7 the results are compared internally and
with results from the literature.
% section 2
\section{Samples and Observations}
% table 1
\begin{table*}
\caption{Details of spectroscopic observing runs.}
\label{spruns}
\begin{center}
\begin{tabular}{llccccc} \hline 
\ & \multicolumn{4}{c}{~}   \\ [-1.5ex]
Run & Dates of Observations & Telescope & Wavelength & $\Delta\lambda$ & 
Res
%\footnote{Instrumental resolution (found from FWHM of arc lines)} 
&
Res
%\footnote{Spectrograph resolution (instrumental dispersion)} 
\\ 
Code & & & Range ($\mbox{\AA}$) & ($\mbox{\AA}$/pix) & 
($\mbox{\AA}$) & (km~s$^{-1}$) \\
\ & \multicolumn{4}{c}{~}   \\ [-1.5ex]
\hline   
\\ [0.5ex]
S1 & 28/06/93 -- 04/07/93 & MDM 2.4m & 4290 -- 6780 & 2.43 & 4.7 & 110 \\ [0.5ex]
S2 & 16/06/94 -- 21/06/94 & MDM 2.4m & 4440 -- 6730 & 2.24 & 3.9 & ~93  \\ [0.5ex]
S3 & 11/09/94 -- 18/09/94 & MDM 2.4m & 4675 -- 7035 & 2.31 & 5.0 & 118 \\ [0.5ex]
S4 & 25/09/94 -- 27/09/94 & MMT 4.4m & 5000 -- 5970 & 0.81 & 1.6 & ~40 \\ [0.5ex]
S5 & 21/10/94 -- 23/10/94 & MDM 2.4m & 4295 -- 7165 & 2.81 & 5.6 & 133 \\ [0.5ex]
S6 & 09/03/95 -- 14/03/95 & MDM 2.4m & 4305 -- 6540 & 2.19 & 4.4 & 103 \\ [0.5ex]
S7 & 29/05/95 & MDM 1.3m & 4000 -- 7100 & 3.04 & 5.6 & 132 \\ [0.5ex]
S8 & 05/06/95 -- 11/06/95 & MDM 2.4m & 4320 -- 6565 & 2.20 & 4.3 & 102 \\ [0.5ex]
S9 & 01/09/95 -- 04/09/95 & MDM 2.4m & 4320 -- 6565 & 2.20 & 4.3 & 102 \\ 
& & & & &  \\ \hline
\ & \multicolumn{4}{c}{~}   \\ 
   \end{tabular} \vspace{-.1in}
\end{center}
\end{table*}
% table 2
\begin{table*}
\caption{CCD and instrument parameters for spectroscopy.}
\label{ccdins}
\begin{center}
    \vspace{.1in}
\begin{tabular}{ll@{~~~}l@{~~~}l@{~~~}l} \hline
\ & \multicolumn{4}{c}{~}   \\ [-1.5ex]
& \multicolumn{4}{c}{Run Code~~~~~~~~~~~~~}  \\ 
\ & \multicolumn{4}{c}{~}  \\ [-2.5ex]
& \multicolumn{4}{c}{---------------------------------------------------------------------------------}  \\ 
\ & \multicolumn{4}{c}{~}  \\ [-2.5ex]
Parameter & S1,~S2,~S3, & S4 & S5 & S7 \\ 
& S6,~S8,~S9 & & & \\ 
\ & \multicolumn{4}{c}{~}   \\ [-1.5ex]
\hline
\\ [0.5ex]   
Telescope & MDM 2.4 m & MMT 4.4 m &MDM 2.4 m&MDM 1.3m\\[0.5ex]
Spectrograph & Mark III&Red Channel&Mark III&Mark III\\[0.5ex]
Grating/grism ($l$/mm) & grism (600) & grating (1200) & grism (600) & grism (600) \\ [0.5ex]
Blaze ($\mbox{\AA}$) & 5800 & 5750 & 5800 & 4600 \\ [0.5ex]
Slit width (arcsec) & 1.68 & 1.00 & 1.68 & 2.17 \\ [0.5ex]
Detector & Tek CCD & Loral CCD & Loral CCD & Loral CCD\\
Format & $1024 \times 1024$ & $800 \times 1200$ & $2048 \times 2048$ & $2048 \times
2048$ \\
Binning & $1 \times 1$ &  $2 \times 1$  & $2 \times 2$ & $2 \times 2$\\[0.5ex]
Readout Noise (e$^{-}$) & 6.0 & 7.0 & 5.4 & 5.4 \\ [0.5ex]
Gain (e$^{-}$/ADU) & 3.45 & 2.60 & 1.94 & 1.94 \\ [0.5ex]
Pixel size ($\mu$m) & 24 & 15 & 30 & 30 \\ [0.5ex]
% This is the pixel size with binning
Spatial scale (arcsec/pixel) & 0.777 & 0.300 &0.971 &1.794 \\ 
% This is the scale in the wavelength direction, with binning and reduction
& & & & \\ \hline 
   \end{tabular} \vspace{-.1in}
   \end{center}
\end{table*}
Sample galaxies were selected from the new Automatic Plate Measuring Facility (APM) 
South Equatorial Strip Catalog, made available by Somak Raychaudhury prior to publication
(Raychaudhury $et$ $al.$ 1999). 
The South Equatorial Strip Catalog ($-17^{\circ}.5 < \delta < +2^{\circ}.5$) is an uncharted
region in the velocity field, because previously no good galaxy catalog existed for this region,
and consequently the peculiar motions of galaxies in this strip had never been mapped. 
Redshifts had been previously measured for only about 20 of the sample galaxies, and velocity
dispersion data existed for only a few of the sample galaxies. 

The sample consists of early-type galaxies in three selected directions: (1) 15h10 -- 16h10; (2)
20h30 -- 21h50; (3) 00h10 -- 01h30. The first region is about 20$^{\circ}$ from the direction
of the LP bulk flow, the second region is almost perpendicular to the first, and the third is on the
opposite side of the sky from the first, close to the direction of the Perseus-Pisces region and the
South Galactic Pole.
The APM South Equatorial Strip Catalog has a magnitude limit of $b_j = 17.0$ mag, which
corresponds to 15.05 mag in Kron-Cousins $R$. All candidate galaxies in the three regions were
examined on the POSS plates and then on CCD images to verify the morphological type. 
The original intention was to observe all galaxies down to this magnitude limit. We have both
photometric and spectroscopic data for 179 of these galaxies, which resulted in a sample of E/S0
galaxies virtually complete to $R$ = 14.0. The completeness drops for fainter magnitudes;
galaxies down to $R$ = 15.0 are included.
New observations were carried out for all 179 galaxies, so that the sample has a fully
independent data set with homogeneous observations, uniform data reduction, and consistent
measurement techniques for all the galaxies.

In addition to the galaxies in the three sample regions, a number of standard galaxies from
previous studies were also observed, for comparison purposes, in order to confirm the accuracy
of the spectroscopic parameters obtained in this work. Comparison galaxies were selected from
the following samples in the literature: Davies $et$ $al.$ (1987); Gonz\'{a}lez (1993); McElroy
(1995); and J$\o$rgensen $et$ $al.$ (1995).
Spectra were also obtained for a sample of 40 galaxies in the Coma cluster, which was used as
the calibration cluster for the FP distance-indicator relation. Galaxies were chosen to be E/S0
using earlier studies ($e.g.$ Lucey $et$ $al.$ 1991; J$\o$rgensen $et$ $al.$ 1993), and most lie
within about $0.5^{\circ}$ of the point midway between NGC~4889 and NGC~4874.

Spectroscopic observations were done at the 2.4~m Hiltner telescope and the 1.3~m
McGraw-Hill telescope of the Michigan-Dartmouth-M.I.T. (MDM) Observatory on Kitt Peak,
Arizona, and also at the 4.4 m Multiple Mirror Telescope\footnote{The Multiple Mirror
Telescope Observatory is operated by the Harvard-Smithsonian Center for Astrophysics jointly
with the University of Arizona.} (MMT) on Mount Hopkins, Arizona. Data were collected
during a series of nine observing runs between June 1993 and September 1995. Over 260 galaxy
spectra were obtained. In total, 38 nights were allocated for this project, and 25 of them were
usable. The main details of the observing runs are summarized in Table~1. (The resolution in
$\mbox{\AA}$ given in Table 1 is the instrumental resolution, found from the FWHM of arc
lines, whereas the resolution in km~s$^{-1}$ is the spectrograph resolution or instrumental
dispersion.)
The spectral range was chosen to cover the Mgb band (around $\lambda_0$ = 5177
$\mbox{\AA}$), the E-band (5270 $\mbox{\AA}$) and the Fe I line (5335 $\mbox{\AA}$).
For some runs the H$\beta$ (4861 $\mbox{\AA}$) and Na D (5895 $\mbox{\AA}$) features
were also included. All observations were made with a long slit and a CCD detector. The setup
and instrument parameters for all runs are shown in Table~2.

All observations on the two telescopes at MDM were made with f/7.5 and  
using the Mark III spectrograph, which consists of a grism, glass optics, and a CCD detector. 
Two different detectors were used with the MDM telescopes: the Tektronix TK1024A
1024$^{2}$ CCD, which is a thinned, back-illuminated CCD with a pixel size of 24 $\mu$m,
and the Loral 2048$^{2}$ CCD, which is a thick, front-illuminated CCD with a 15 $\mu$m
pixel size. Both CCDs are good cosmetically and have low readout noise.
For the observations at the MMT we used the Red Channel: a spectrograph consisting of a
collimator, a folding flat, a grating, and a CCD detector. The spectrograph was used in the
high-throughput long-slit mode with a slit 180 arcsec long. The detector used at the MMT was
a Loral $800 \times 1200$ CCD binned by 2 pixels in the narrower spatial direction
(perpendicular to the dispersion). The CCD is very good cosmetically, with only a few traps.

The slit was usually oriented with the long axis running North-South. The instrumental
resolution FWHM (in pixels and in $\mbox{\AA}$) was determined by fitting a Gaussian to
measure the widths of lines in calibration lamp spectra and of night-sky lines.
Each galaxy was observed with a sufficiently long exposure time to ensure a high enough
signal-to-noise ratio ($S/N \sim$ 20) to enable the central velocity dispersion, as well as the
redshift of the galaxy, to be determined accurately from the spectrum. At the MDM 2.4 m
telescope, the average integration time needed was 2400~s; the integration times for individual
galaxies ranged from 900~s to $2 \times 3600$~s according to the magnitude and surface
brightness of the galaxy and the observing conditions. At the MMT, the average integration time
per galaxy was 1200~s.
A total of 263 spectra were obtained, of 238 galaxies. For some sample galaxies, more than one
spectrum was obtained; an observation was repeated in some cases to provide a way to check
the accuracy.
Spectra were also obtained for KPNO IIDS spectrophotometric flux standard stars (Strom 1979),
for use in flux calibration of the spectra. The exposure time was usually 360~s, and one flux
standard star was observed per night.

% table 3
\begin{table}
\begin{flushleft}
\caption{Radial velocity standard stars and template stars.}
\label{templates}
    \vspace{.1in}
\begin{tabular}{l@{~~~~~~~~~}l@{~~~~~~}rr} \hline 
\ & \multicolumn{3}{c}{~}   \\ [-1.5ex]
Star & Spectral & $S/N$ & Radial Velocity\\
 & Type & per $\mbox{\AA}$ & (km~s$^{-1}$)\\
\ & \multicolumn{3}{c}{~}   \\ [-1.5ex]
\hline   
\\ [0.5ex]
HD ~4388     &           K3 III  & 62  & $-28.3$  \\ [0.5ex]  
HD 12029     &           K2 III  & 350 & +38.6    \\ [0.5ex]    
HD 20893     &           K3 III  & 110 &  +5.9    \\ [0.5ex]    
HD 22072     &           DG7     & 120 & +12.2    \\ [0.5ex]    
HD 36003     &           K0      & 106 & $-58.2$    \\ [0.5ex]  
HD 38751     &           G8 III  & 220 & +15.7    \\ [0.5ex]    
HD 51440     &           K2 III  & 117 & +27.1    \\ [0.5ex]    
HD 64606     &           G5      & 137 & +93.8   \\ [0.5ex]     
HD 65934     &           K0      & 130 & +35.0    \\ [0.5ex]    
HD 72324     &           G9 III  & 68  & +75.2   \\ [0.5ex]     
HD 73665     &           K0 III  & 120 &  +36.9  \\ [0.5ex]     
HD 74377     &           K0      & 84  &  $-25.4$  \\[0.5ex]    
HD 90861     &           K2 III  & 140 &  +36.3   \\[0.5ex]     
HD 132737    &           K0 III  & 110 & $-24.1$   \\ [0.5ex]   
HD 165195    &           G5      & 50  & $-0.2$    \\ [0.5ex]   
HD 171232    &           G8 III  & 142 & $-35.9$   \\ [0.5ex]   
HD 172401    &           ~~~--     & 102 & $-73.0$ \\[0.5ex]    
HD 194071    &           G8 III  & 127 & $-9.8$    \\ [0.5ex]   
HD 213947    &           K4 III  & 280 & +16.7   \\ [0.5ex]    
HD 223094    &           K5 III  & 375 & +19.6   \\ 
& & &  \\ \hline     
\ & \multicolumn{3}{c}{~}   \\ 
   \end{tabular} \vspace{-.1in}
\end{flushleft}
\end{table}
In addition, several stars with known radial velocities were observed during each run. (Values
of radial velocity for the standard stars were taken from Wilson 1953, Evans 1970, Abt \& Biggs
1972, and Barbier-Brossat \& Petit 1987.)
These radial velocity standard stars were chosen to be of luminosity class III and in the range of
spectral types G5 -- K5. Most were K giants fainter than 7th magnitude. The observed stars are
listed in Table~3.
%In addition, a few dwarfs of F and G types were observed.
The same set of stars served as spectral templates for the determination of the velocity
dispersions of the galaxies and as radial velocity standards for finding the galaxy redshifts. 
All the standard stars were trailed up and down the slit so that the slit illumination would be
uniform and more similar to that of a galaxy. The exposure time was normally 420~s, 
which usually resulted in a very high $S/N$ ratio.

Before and after the spectrum of each galaxy or standard star, a comparison spectrum was taken
for use during wavelength calibrations. At MDM, Hg + Ne and Ne lamps were exposed for
0.2~s, and at the MMT He + Ne + Cu + Ar lamps were exposed for 120~s. Each night a series
of bias frames was taken, as well as spectral flat field frames using an internal continuum flat
lamp. Exposure times for flat fields were 30~s at MDM and 2~s at the MMT.
% section 3
\section{Preliminary Data Reduction Procedures}
Packages in the image processing system IRAF\footnote{IRAF is distributed by National Optical
Astronomy Observatories, which is operated by the Association of Universities for Research in
Astronomy, Inc., under cooperative agreement with the National Science Foundation, USA.}
(Tody 1986) were used for the basic reduction of the spectra. The main preliminary reduction
steps, which are described in more detail below, are the following:\
(i)~bias subtraction and dividing the data by a flat field;\ 
(ii)~mapping the wavelength as a function of row and column by using the comparison
exposures;\
(iii)~subtracting the sky spectrum;\
(iv)~removing the cosmic-ray hits from the spectra; and\ 
(v)~extracting one-dimensional spectra from a sum over the aperture.

The bias level for each frame was found from the overscan region. The bias frames were
averaged and the residual bias level subtracted pixel-by-pixel from each image.
>From dark frames with exposures of 2400~s, a level always much less than one count was found.
Since the dark current is flat, no dark count corrections were made.
The flat field frames were combined to their median value, 
by using {\tt noao.imred.ccdred.flatcombine}. This result was then normalized by fitting a cubic
spline to the continuum in the wavelength direction and dividing the flat field by this fit to obtain
the response function. The rms variation in the resulting flattened response frames was typically
less than 0.5\,\%. Each galaxy or star spectral frame was then divided by the response function.

The next step was to transform from pixel coordinates to a two-dimensional spatial scale with
wavelength coordinates along the dispersion axis of the CCD image. Tasks in the package {\tt
noao.twodspec.longslit} were used for this. By means of 25 -- 50 identified arc lines in the
comparison lamp spectra, a polynomial was fitted to the wavelength solution, with an rms
residual in the coordinate fit of 0.1 -- 0.2~$\mbox{\AA}$. The two-dimensional spectra of stars
and galaxies were calibrated in wavelength using the comparison lamp spectra observed before
and after each object, and thus transformed to a linear wavelength scale.

The sky was fitted interactively using {\tt background}, with the sky level found from
unoccupied regions of the slit, and the sky background was subtracted. This worked well, since
the galaxy occupied a relatively small part of the slit. Cosmic rays were removed by rejection
using {\tt images.lineclean}, with care being taken that absorption lines were unaffected.
The final one-dimensional spectra were extracted by summing over an aperture covering the
entire visible galaxy. This included tracing the mapping of the slit position across the CCD as
a function of row or column. The package {\tt kpnoslit.apall} was used interactively to fit the
traced positions of the apertures.  For all runs the trace was found to vary by at most 4 pixels
across the CCD, in 1023 pixels. The typical width of the galaxy spectrum was about 20 pixels,
so the maximum misalignment resulting from the tilt of the spectrum would be 0.1 pixels.
% figure 1
\begin{figure} 
%\picplace{7cm}
%\centerline{
\psfig{file=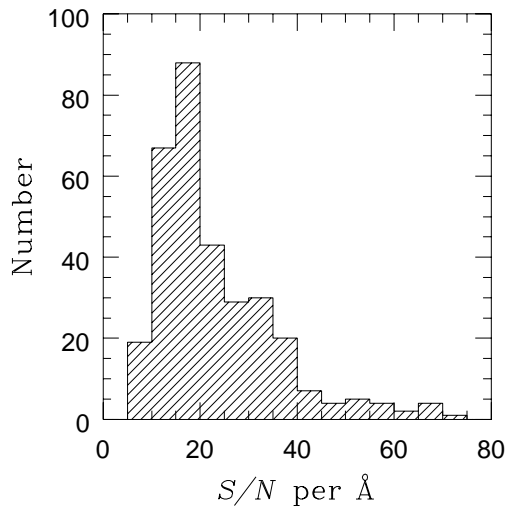,height=70mm}
%}
\caption{The distribution of spectra as a function of the mean $S/N$ per $\mbox{\AA}$
measured at the wavelength of Mgb.}
\label{snhist}
\end{figure}

The summed one-dimensional spectra typically have 
1000 -- 20~000 counts at 5200 $\mbox{\AA}$, near the wavelength of Mgb. 
%The galaxy-to-sky ratio is highest at longer wavelengths. 
The mean value of the $S/N$ per $\mbox{\AA}$ was found for each spectrum, of galaxies and
standard stars. The $S/N$ was calculated from the mean number of photon counts in the
spectrum in the wavelength range of the continuum bands of the Mgb spectral feature, also
including in the noise the contribution of the readout noise and the effect of subtracting the sky
spectrum. For most spectra the resulting value of $S/N$ per $\mbox{\AA}$ is in the range of 15
-- 40. A histogram of the frequency distribution of all the spectra with respect to $S/N$ is shown
in Fig.~1. The mean $S/N$ for all spectra is 23.0. For sample galaxies it is 21.4, for Coma
galaxies 27.3, and for standard galaxies 26.8. The $S/N$ of a typical stellar observation is 100
-- 200, which means that the stellar spectra can be considered to be noiseless.

At this stage the one-dimensional extracted spectra were inspected. Usually the sky subtraction
was reasonably accurate, but the spectra were checked for night-sky emission lines, particularly
those of [O I] at 5577 $\mbox{\AA}$ and Hg at 5461 $\mbox{\AA}$, and if necessary these
lines were removed by hand in cases of imperfect sky subtraction.
There were sometimes cosmic rays which had not been removed completely, and these were also
removed by hand. Three of the observed galaxies had spectra which were found to contain strong
emission lines, and these galaxies were removed from the sample.
% section 4
\section{Determination of Redshifts and Velocity Dispersions}
The accurate determination of the line-of-sight central velocity dispersions of the galaxies is
critical to FP analysis. A number of different methods exist for determining the velocity
dispersion $\sigma$ and the radial velocity $cz$. These include the Fourier quotient method
(Sargent $et$ $al.$ 1977), the Fourier cross-correlation method (Tonry \& Davis 1979), the
Fourier difference method (Dressler 1979), and the Fourier fitting method (Franx $et$ $al.$
1989).

For this study we used the Fourier cross-correlation method, as implemented in the IRAF
package {\tt rv.fxcor} which is based on the method of Tonry \& Davis (1979). 
The spectra of the galaxy and the stellar template are cross-correlated in Fourier space, and the
resultant maximum peak is fitted by a smooth symmetric function. The width and pixel shift of
the peak are measures of $\sigma$ and the galaxy redshift (in km~s$^{-1}$), found by
comparison with the known radial velocity of the template. An indicator of the accuracy of the
resulting value of $\sigma$ is the $r$ value (Tonry \& Davis 1979).

All spectra were rebinned, using {\tt onedspec.dispcor}, to linear logarithmic wavelength
coordinates. 
Total flux was conserved, and the same dispersion parameters were used for all spectra from all
runs, resulting in spectra with logarithmic wavelength bins of $\Delta \ln \lambda = 6.46 \times
10^{-5}$.
A cubic spline was fitted to the continuum for all spectra  (template stars as well as galaxies). 
This fit was subtracted from the spectrum to flatten it; the resulting spectrum has zero mean in
the continuum.
The spectra were then Fourier filtered before the correlation. Data points outside the selected
sample region were zeroed, and the ends of the region (12.5\,\% on each end of the spectrum)
were apodized with a cosine bell. A ramp function was used as the filter. The parameters of the
filter were adjusted to find the best combination. 
After the galaxy and template spectra had been thus prepared, the two sets of spectra were
cross-correlated. 
In most cases the best region for cross-correlation was found to be 4900 -- 5800 $\mbox{\AA}$.
This choice excludes the H$\beta$ and Na D lines.

The galaxy spectra were Fourier cross-correlated in {\tt fxcor} against each standard star in turn. 
The observed FWHM of the cross-correlation peak was transformed into a value for $\sigma$
by direct calibration with broadened template spectra, using the preocedure outlined by Baggley
(1996).
The spectrum of each template star was convolved with Gaussians of various known widths in
the range 0 -- 700~km~s$^{-1}$, and the resulting broadened spectra were run in {\tt fxcor}
(with the same parameters) against the original template spectrum, giving the FWHM of the
cross-correlation peak in each case. A calibration curve of this FWHM width versus the
broadening $\sigma$ for Gaussians of different widths was produced for each template star
observation, by linear interpolation between the FWHM values from {\tt fxcor}.
The galaxy FWHM values were then converted into values for $\sigma$ for the galaxy by
reading off the calibration for that particular template star. An example of a calibration curve
is shown in Fig.~2, for HD 194071, observed during run S8. In this case the resolution was
approximately 100~km~s$^{-1}$, and it can be seen from the figure that below this value of
$\sigma$ it is more difficult to find an accurate determination of velocity dispersion.

For every galaxy there is a set of different values for $\sigma$ and $cz$, each pair of values the
result of running the galaxy against a different stellar template spectrum. 
The values obtained by Fourier cross-correlation show small systematic differences depending
on which stellar template is used. Template stars from all runs were used, a total of 45
observations of 20 different stars (as listed in Table~3). 
The rms difference between the estimates for $cz$ and $\sigma$ from different template stars
was typically $<$ 1\,\% in $cz$ and $\sim$ 4\% in $\sigma$.

% figure 2
\begin{figure} 
%\picplace{8cm}
%\centerline{
\psfig{file=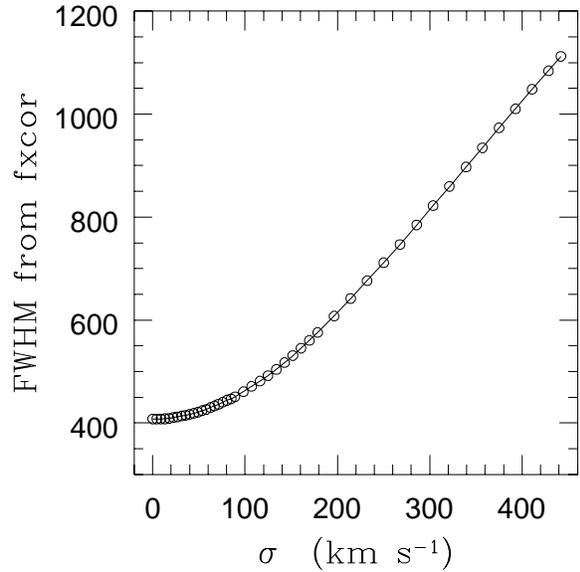,height=80mm}
%}
\caption{ The calibration curve for an observation of the standard star HD 194071.
The points represent the FWHM from {\tt fxcor} for the broadened template spectra compared
with the unbroadened original template spectrum. The curve is found from linear interpolation
between the points.}
\label{calcurve}
\end{figure}
% figure 3
\begin{figure} 
%\picplace{8.8cm}
%\centerline{
\psfig{file=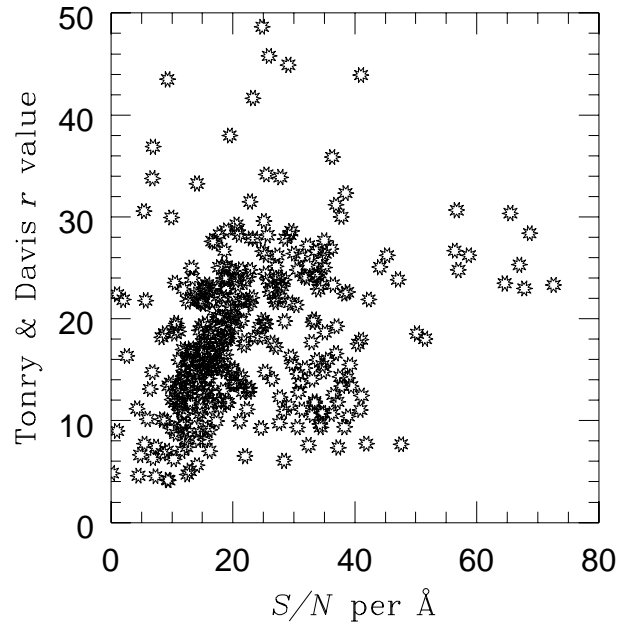,height=88mm}
%}
\caption{A comparison of the $S/N$ per $\mbox{\AA}$ of the spectra with the value of the
Tonry \& Davis $r$ parameter from the cross-correlation, which is a measure of the ratio
between the the peak height and the noise, and is therefore an indicator of the accuracy of the
results.}
\label{sntdr}
\end{figure}
 In Fig.~3 the mean $r$ value from the cross-correlation analysis of a spectrum is shown plotted
against the $S/N$ of that spectrum. From this plot it can be seen that the scatter is larger than
would be expected if the $S/N$ were the only factor affecting the $r$ value.
It could be that $r$ is also sensitive to a mismatch between the features of the galaxy and those
of some of the template spectra. 
Dalle Ore $et$ $al.$ (1991) found no systematic variation of the width of the cross-correlation
curve with spectral type, and showed that errors in the velocity dispersion owing to spectral type
mismatch are negligibly small. J$\o$rgensen $et$ $al.$ (1995) found that template stars of the
spectral type G8 -- K3 result in significantly better fits than stars of types K4 -- K5. For our
sample data, we do not find this to be true in general. For a particular galaxy, certain templates
work better than others, but we find that the best set of templates is different for each galaxy.
We therefore combined the results for the velocity dispersion for each galaxy in such a way as
to minimize the effect of the template mismatch problem. First, the mean of all the results from
different templates was taken. Then the 2$\sigma$ outliers were excluded and the mean was
taken again, to give the final result. The redshift for each galaxy was estimated by calculating
the mean observed heliocentric velocity from all the templates. For both $\sigma$ and $cz$, the
standard deviations from the means were calculated.

To check the zero-point for the redshift determinations, the spectrum of each radial velocity
standard star was cross-correlated with the spectra of all the other standard stars, resulting in
estimates of relative velocity.
The radial velocity standards were run against each other in {\tt fxcor} using the same scripts
with the same parameters as used for the galaxies. For each star the mean of the estimates of
heliocentric redshift was found, and compared with the known value of radial velocity for that
star.
The rms difference between the mean estimated value and the known value was about 20
km~s$^{-1}$, which is therefore the accuracy of the zero-point of the radial velocities we
determined. 

Since a variation was observed among the separate determinations from each of the template
spectra, estimates of the systematic errors for both redshift and velocity dispersion have been
obtained from the rms scatter of the results from different templates. These error estimates are
listed together with the results in Tables~5, 6, and 7.
% section 5
\section{Corrections to Redshifts and Velocity Dispersions}
% figure 4
\begin{figure*} 
%\picplace{8cm}
%\centerline{
\psfig{file=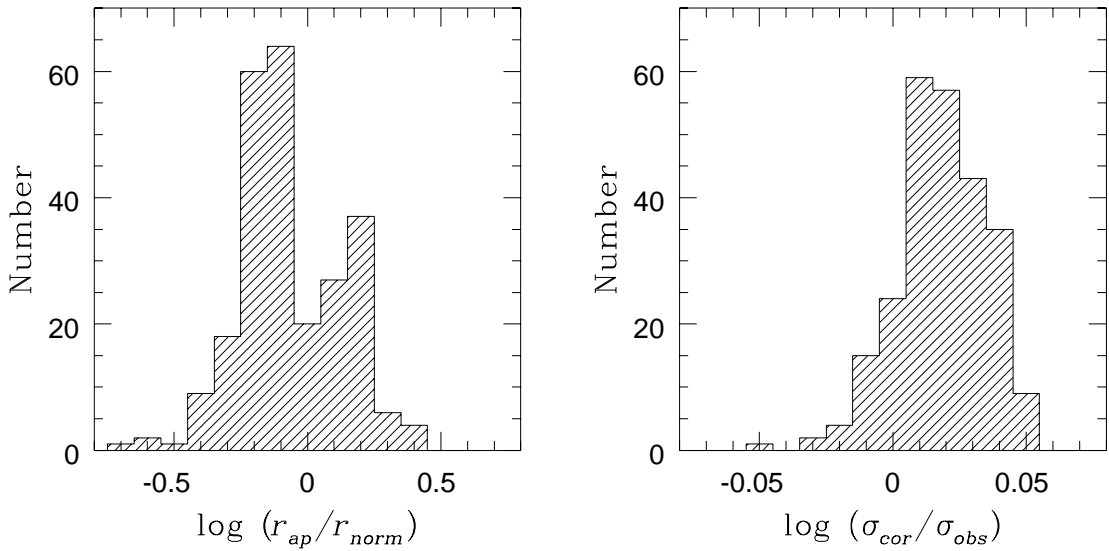,height=80mm}
%}
\caption{Histograms for the galaxies in the three sample regions and in the Coma cluster. {\it
Left:} A histogram of the aperture sizes of the sample galaxies relative to the normalising
aperture radius, with $r_{ap}$ in units of kpc at the distance (redshift) of the galaxy, and 2
$r_{norm}$ = 1.19 $h^{-1}$ kpc, equivalent to 3.4 arcsec for a galaxy at the distance of the
Coma cluster ($h$ = 1 was used). \,
{\it Right:} A histogram of the sizes of the aperture corrections applied to the sample galaxies;
$\sigma_{obs}$ is the raw observed value and $\sigma_{cor}$ is the aperture-corrected value
of the velocity dispersion.}
\label{ratios}
\end{figure*}
\subsection{Heliocentric Correction to Redshifts}
Redshifts were corrected for the radial velocity of the template star and also corrected to the
heliocentric system to take into account the motion of the Earth relative to the observed galaxy
and the template star. This correction is included in {\tt fxcor}; the result in the heliocentric
frame is denoted $v_{hel}$. 
\subsection{Aperture Correction to Velocity Dispersions}
In E and S0 galaxies there are radial gradients in the velocity dispersion, with a higher velocity
dispersion in the center of the galaxy than in the outer regions (Davies \& Birkinshaw 1988;
Franx $et$ $al.$ 1989; Davies $et$ $al.$ 1993). For this reason the derived "central" velocity
dispersion parameter depends on the distance of the galaxy and on the size of the aperture used
for the observation of the spectrum. It is therefore necessary to apply an aperture correction to
transform the observed parameters so that they are independent of distance and of the telescope
used.

The measured value of velocity dispersion depends on the velocity dispersion profile in the
galaxy. Since the profile is not known for each galaxy, a general form must be assumed.
J$\o$rgensen $et$ $al.$ (1995) established an aperture correction from kinematic models based
on the available literature data. They used the models to derive the equivalent circular aperture
for each rectangular aperture, and adopted a power law as the aperture correction to $\sigma$.
The radius $r_{ap}$ (in arcsec) of the equivalent circular aperture is found from $r_{ap} =
1.025 \sqrt(wl/\pi)$, where $w$ and $l$ are the width and length of a rectangular aperture (slit). 
J$\o$rgensen $et$ $al.$ (1995) correct the velocity dispersion to an aperture with a standard
physical size, and use a value for the normalising aperture size of 2 $r_{norm}$ = 1.19 $h^{-1}$
kpc ($h \equiv H_0$ / 100 km~s$^{-1}$ Mpc$^{-1}$.) This means that the velocity dispersion
values are normalized to a velocity dispersion measured with an aperture of diameter 1.19
$h^{-1}$ kpc, which is equivalent to 3.4 arcsec for a galaxy at the distance of the Coma cluster.
 
Baggley (1996) has shown that it is necessary to also take the effective radius $r_e$ into account
in the aperture correction, since the observed velocity dispersion of a galaxy depends on $r_e$
as well. For two galaxies of different sizes at the same distance observed through the same
aperture, the slit will cover more of the smaller galaxy and a different part of the galaxy profile
will therefore be sampled; this dependence of $\sigma$ on $r_e$ must be removed to ensure that
the velocity dispersion is truly a distance-independent quantity. We used Baggley's formula for
the aperture correction; it is a generalisation of the formula of J$\o$rgensen $et$ $al.$ (1995)
to take $r_e$ into account:
\begin{equation}
\label{apcorr}
\,\log \frac{\sigma_{cor}}{\sigma_{obs}} = 0.038 \, \log \left( \frac{r_{ap}}{r_{norm}} \,
\frac{cz_{gal}}{cz_{Coma}} \, \frac{r_e^{norm}}{r_e^{gal}}\right),
\end{equation}
where: 
$\sigma_{obs}$ is the value of the velocity dispersion found from observation through an
aperture equivalent to $r_{ap}$;
$\sigma_{cor}$ is the velocity dispersion value corrected to the adopted normalising aperture
size $r_{norm}$; 
$cz_{gal}$ is the redshift of the galaxy; $cz_{Coma}$ is the redshift of the Coma cluster;
$r_e^{gal}$ is the effective radius of the galaxy; and 
$r_e^{norm}$ is the normalising effective radius, which is taken to be 20 arcsec, following
Baggley (1996) -- this is the mean $r_e$ of the galaxies in the sample of J$\o$rgensen $et$ $al.$
(1995), which was used to derive the correction. 
This means that there will be no aperture correction for a galaxy with $r_e$ = 20 arcsec, at the
distance of the Coma cluster, observed through an aperture with an equivalent radius of 1.7
arcsec.

Eq.~1 was used to calculate the corrections.
The slit widths in the various instrumental setups are as shown in Table~2.
For the length of the slit in the aperture, what is important is the extent of the galaxy along the
slit, i.e. the length from which the one-dimensional spectrum was extracted; the spectrum was
extracted out to the point where the luminosity had fallen to 10\,\% of its peak value.
The $r_e$ value for each galaxy was taken from the results of the fitting and seeing-correction
programs applied to the photometric data, as described in M\"uller (1997) and
M\"uller $et$ $al.$ (1999).
The value of the heliocentric redshift $cz_{Coma}$ was 
taken to be 6917 km~s$^{-1}$ (Zabludoff $et$ $al.$ 1993), 
and 
%$H_0$ = 100 km~s$^{-1}$ Mpc$^{-1}$ ($i.e.$ 
$h$ = 1
%) 
was used in the conversion of km~s$^{-1}$ to kpc. % 7219 in CMB frame

The contribution to the aperture correction from the effect of $r_e$ is more important than those
for different slit sizes and different galaxy distances.
>From Eq.~1 it can be seen that the correction is negative for large or nearby objects (this is the
case for the standard galaxies) and positive for galaxies with $r_e$ less than 20 arcsec, which
are smaller than or more distant than a 20 arcsec galaxy at the distance of Coma.
The histogram of the applied aperture corrections in Fig.~4 shows that for the sample galaxies
the correction is positive in most cases.
% section 6
\section{Results}
\subsection{Quality Ratings for Spectroscopic Parameters}
% table 4
\begin{table*}
\begin{center}
\caption{Percentage distribution of quality ratings.}
\label{percent}
\vspace{.1in}
\begin{tabular}{lccccc} \hline 
\ & \multicolumn{2}{c}{~}   \\ [-0.5ex]
Quality Rating & \multicolumn{5}{c}{Percentage of Sample Galaxies} \\ [-1.5ex]
& & \\
& Standard galaxies & Region~1 & Region~2 & Region~3 & Coma galaxies \\
& & \\ [-0.5ex]
\hline  
& & \\
$Q_{spec}$ = 1 & ~72 & ~24 & ~62 & ~63 & ~43  \\ [0.5ex]  
$Q_{spec}$ = 2 & ~28 & ~26 & ~26 & ~28 & ~32  \\ [0.5ex]  
$Q_{spec}$ = 3 & ~~0 & ~46 & ~10 & ~~9 & ~25 \\ [0.5ex]   
$Q_{spec}$ = 4 & ~~0 & ~~4 & ~~1 & ~~0 & ~~0 \\   
& &  \\  
\hline
\ & \multicolumn{2}{c}{~}   \\ 
   \end{tabular} \vspace{-.1in}
\end{center}
\end{table*}
% figure 5
\begin{figure*} 
%\picplace{10cm}
%\centerline{
\psfig{file=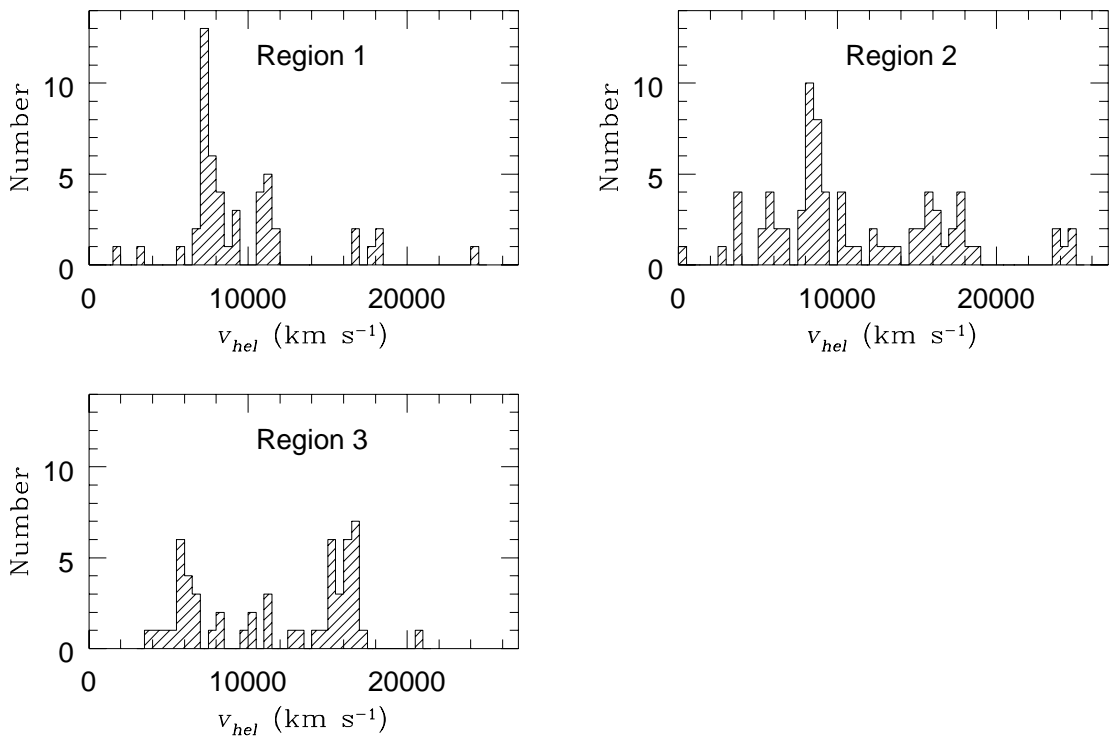,height=100mm}
%}
\caption{Histograms of heliocentric radial velocity for observed galaxies in the three sample
regions. Bin width 500 km~s$^{-1}$.}
\label{czhistsamp}
\end{figure*}
% figure 6
\begin{figure*} 
%\picplace{5cm}
%\centerline{
\psfig{file=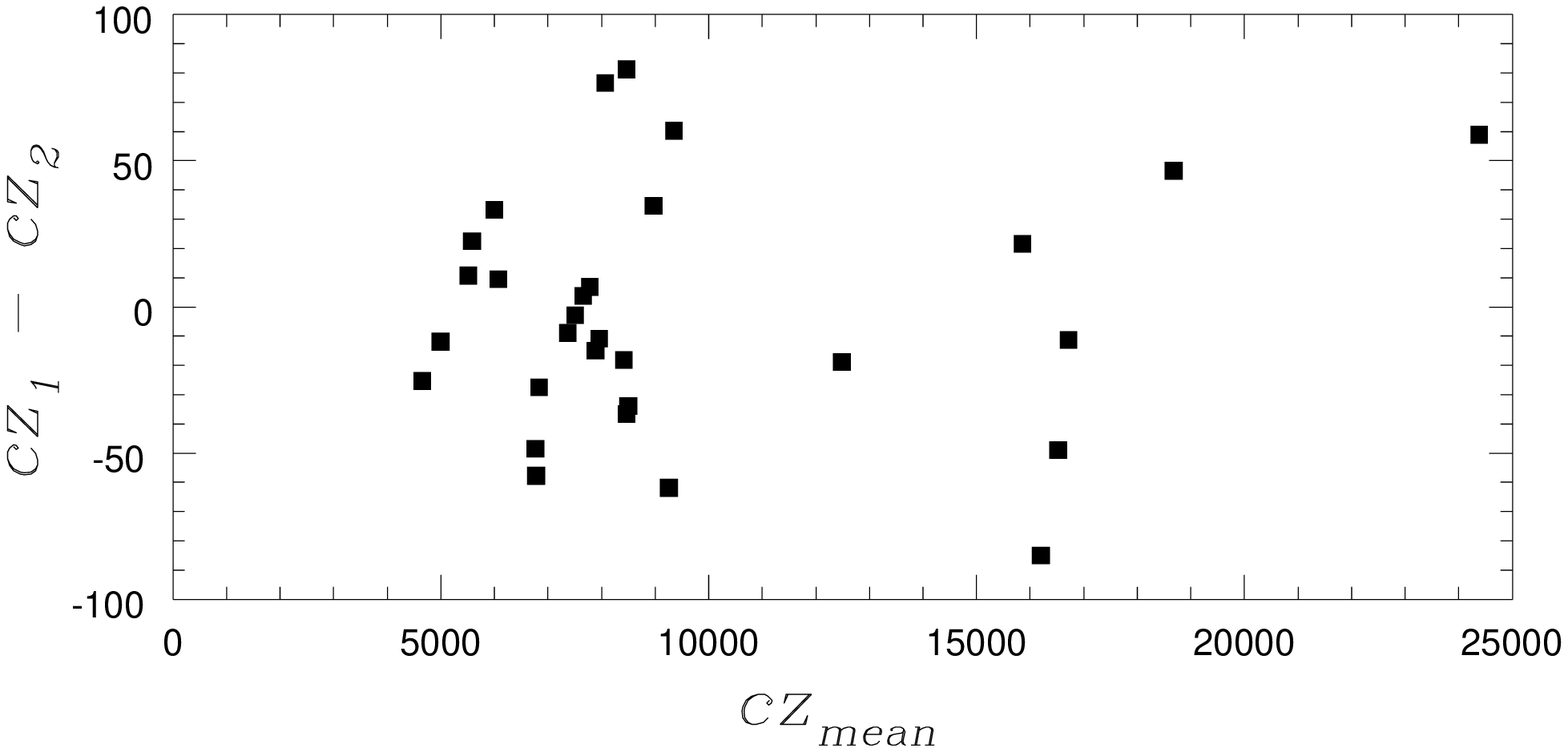,height=55mm}
%}
\caption{Comparison of results for heliocentric radial velocity values determined from repeat
observations.}
\label{intred}
\end{figure*}
A quality rating was assigned to the results from each spectrum, on the basis of the
cross-correlation results, and also after visual inspections of the spectra.
Quality ratings were made as follows: $Q$ = 1 for excellent spectra ($S/N$ per $\mbox{\AA}
>$ 20; mean $S/N \sim$ 30);
$Q$ = 2 for good spectra (mean $S/N$ $\sim$ 16); $Q$ = 3 for acceptable spectra (mean $S/N$
$\sim$ 11); and $Q$ = 4 for poor spectra ($S/N <$ 10; mean $S/N$ $\sim$ 4).
The mean Tonry \& Davis $r$ values for the four quality ratings are $\sim$ 23, $\sim$ 17,
$\sim$ 13, and $\sim$ 7 for ratings of $Q_{spec}$ =  1, 2, 3, and 4 respectively.
The percentage distribution of the quality ratings is shown in Table~4. For the spectra in the
three sample regions, 78\,\% of the spectra are of quality 1 or 2.
%There were 183, 72, 64, and 4 galaxies in the quality classes 1, 2, 3, and 4 respectively.

The $S/N$ required to obtain a good measurement of radial velocity is lower than the S/N
required to obtain a reliable velocity dispersion. For the galaxies with $Q$ = 4 the signal was not
strong enough to enable a reliable determination of the velocity dispersion. This was the case
for galaxies of low surface brightness for which it was difficult to obtain spectra with a high
enough $S/N$ ratio.  From these weak spectra, only redshifts were found, and these galaxies
were not included in the sample used for distance determination.
Below a certain limit, the values of velocity dispersion are less reliable. In the final sample we
used only galaxies with high enough $\sigma$ values to be determined accurately.
The lower limit on usable $\sigma$ values depends on the instrumental resolution. For the
spectra taken at MDM, with a dispersion of $\sim 2.3 \mbox{\AA}$/pix and an instrumental
resolution of $\sim 130$~km~s$^{-1}$, only velocity dispersions which are larger than
130~km~s$^{-1}$ can be measured reliably. For the spectra from the MMT, for which the
dispersion was $\sim 0.80 \mbox{\AA}$/pix and the instrumental resolution was $\sim
40$~km~s$^{-1}$, the limit is lower.
%Also, Nyquist frequency criterion satisfied.
% figure 7
\begin{figure*} 
%\picplace{5cm}
%\centerline{
\psfig{file=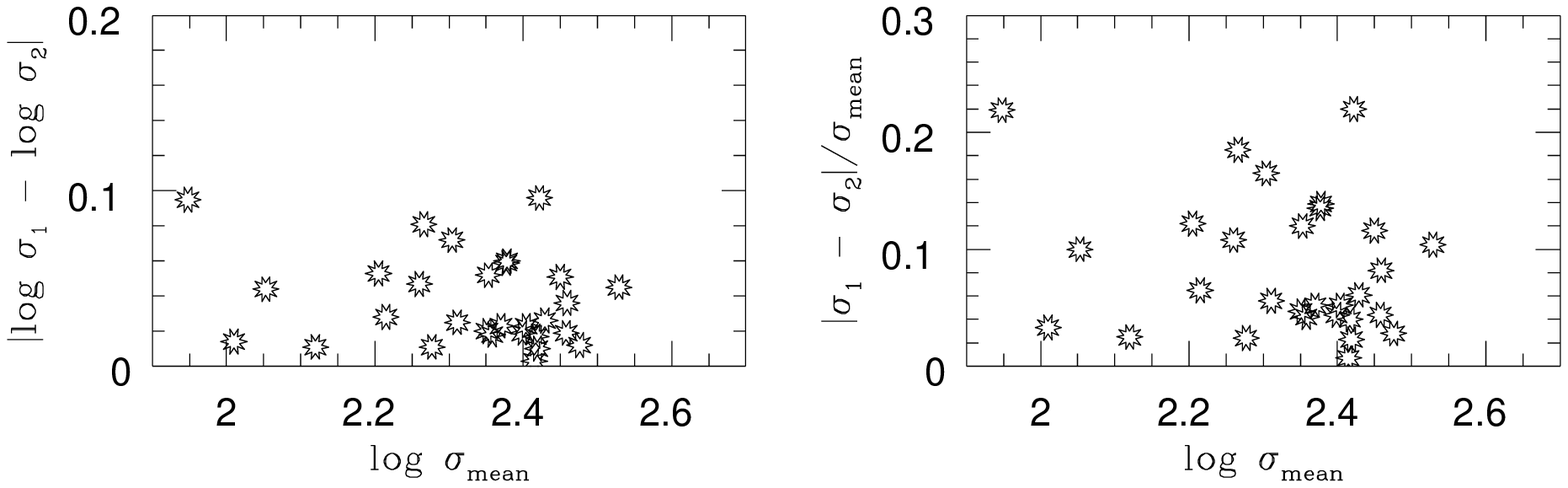,height=50mm}
%}
\caption{Comparison of results for velocity dispersions 
measured from repeat observations.}
\label{intsig}
\end{figure*}
% figure 8
\begin{figure*} 
%\picplace{6cm}
%\centerline{
\psfig{file=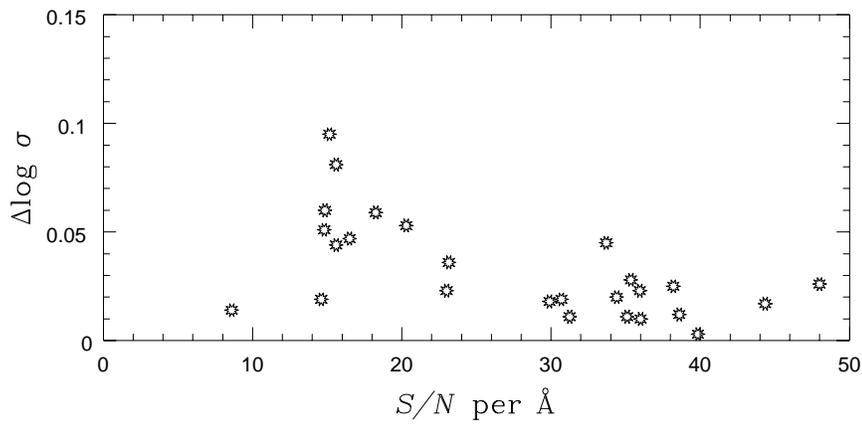,height=60mm}
%}
\caption{The difference between results for velocity dispersions measured from repeat
observations, plotted against the mean $S/N$ per $\mbox{\AA}$ of each pair of spectra.}
\label{sigsn}
\end{figure*}
% table 5
\begin{table*}
\begin{center}
\caption {Spectroscopic parameters for the sample galaxies.}
\label{sampspec}
\vspace{.1in}
\begin{tabular}{lrrrrrrrrc} \hline 
\ & \multicolumn{9}{c}{~}   \\ [-1.5ex]
Galaxy & Right Ascension & Declination & $b_J$ & $d$ &
$v_{hel}$ & err$_{v}$ & $\sigma$ & err$_\sigma$ & $Q_{spec}$\\
\ & \multicolumn{9}{c}{~}   \\ [-1.5ex]
\hline   
& & & \\ [-1.5ex]
{\bf Region~1} &&&&&&&&& \\[0.75ex]
87101144 &16 01 52.14&$-01$ 59 05.0&13.9&1.3&9259&26&305& 4 &1\\[0.5ex]
65301306 &15 16 17.18&$-15$ 48 54.6&14.3&1.2&7296&17&270&16 &1\\[0.5ex]
65300653 &15 24 14.86&$-12$ 48 29.1&14.3&1.2&7351&15&232&21 &1\\[0.5ex]
72501710 &15 17 51.60&$-08$ 21 00.2&14.4&1.3&7573&22&149&18 &3\\[0.5ex]
79800188 &15 47 10.58&$-07$ 19 44.2&14.4&1.3&8032&28&208 &3 &1\\[0.5ex]
65301588 &15 12 23.45&$-14$ 25 57.7&14.5&1.1&7419&18&286 &5 &3\\[0.5ex]
65301596 &15 12 14.43&$-14$ 23 39.5&14.5&1.0&7475&18&168 &9 &1\\[0.5ex]
87001759 &15 36 17.64&$-01$ 34 19.3&14.6&1.3&3439&4 &154 &3 &1\\[0.5ex]
65401564 &15 41 57.11&$-15$ 18 30.0&14.8&1.1&7320&19&209&15 &1\\[0.5ex]
72700354 &16 07 48.41&$-11$ 43 45.9&14.8&0.9&7964&17&109&25 &3\\[0.5ex]
79800136 &15 48 02.65&$-07$ 24 36.0&14.9&0.8&7742&10&231&13 &3\\[0.5ex]
87101109E&16 01 53.82&$-01$ 36 41.8& -- & -- &9069&21&313 &1 &2\\[0.5ex]
65402690 &15 34 32.77&$-16$ 25 59.2&15.0&1.0&7196&18&291&19 &1\\[0.5ex]
CGCG2158 &15 13 02.87&+02 25 56.5&15.0&0.6&1849&22&63 &17   &1\\[0.5ex]
72601433 &15 31 40.06&$-08$ 44 21.2&15.0&0.8&7129&17&265&13 &1\\[0.5ex]
79701386 &15 20 15.85&$-07$ 09 10.5&15.1&1.2&11730&17&239&5 &2\\[0.5ex]
65402668 &15 34 40.55&$-16$ 25 36.1&15.1&0.7&6798&12&147&13 &1\\[0.5ex]
79702357 &15 12 55.64&$-05$ 35 58.1&15.1&1.2&11404&29&228&2 &2\\[0.5ex]
79801072 &15 30 47.53&$-04$ 55 45.2&15.1&0.7&8000&15&171&11 &3\\[0.5ex]
87000696 &15 44 59.18&$-01$ 23 30.8&15.1&0.7&8972&20&276&17 &2\\[0.5ex]
65401595 &15 41 43.20&$-15$ 15 44.8&15.1&0.8&7439&14&175 &9 &2\\[0.5ex]
79700187 &15 29 43.12&$-06$ 35 46.2&15.2&0.9&11112&8&240 &8 &3\\[0.5ex]
65403060 &15 31 44.41&$-16$ 27 53.1&15.2&1.1&7403&15&106 &8 &3\\[0.5ex]
65300973 &15 20 29.81&$-15$ 49 34.3&15.3&0.8&7109&19&130 &7 &3\\[0.5ex]
65501546 &15 58 50.53&$-15$ 22 19.0&15.3&0.9&7510&11&171&10 &3\\[0.5ex]
65402909 &15 32 54.07&$-16$ 37 02.4&15.3&0.9&6849&13&142&15 &3\\[0.5ex]
87002300 &15 30 24.64&+00 36 31.6&15.3&0.8&10980&18 &248&17 &3\\[0.5ex]
65401438 &15 42 48.68&$-16$ 52 32.8&15.4&0.8&11342&28&249&11&2\\[0.5ex]
65300169 &15 29 48.37&$-17$ 26 03.3&15.4&0.7&7050&6 &131 &8 &1\\[0.5ex]
65401086 &15 45 06.95&$-13$ 14 13.8&15.4&0.9&11275&13&184&14&3\\[0.5ex]
65402811 &15 33 40.66&$-16$ 06 32.6&15.5&0.6&7855 &10&206&8 &2\\[0.5ex]
72600774 &15 40 56.84&$-08$ 59 11.0&15.5&0.6&17737&22&316&9 &3\\[0.5ex]
87002308 &15 30 19.55&+00 38 12.6&15.6&0.7&11781 &22&323&15 &2\\[0.5ex]
79901311 &15 57 33.40&$-06$ 46 40.9&15.6&0.7&9146&18&98 &12 &3\\[0.5ex]
79800922 &15 33 30.16&$-05$ 11 58.9&15.6&0.6&7077 &28&127&5 &3\\[0.5ex]
65301134 &15 18 21.41&$-13$ 31 45.3&15.6&0.6&7263 &13&163&15&2\\[0.5ex]
72502368 &15 12 33.13&$-12$ 13 26.2&15.7&0.7&7634 &36&82 &4 &3\\[0.5ex]
65502352 &15 52 05.13&$-17$ 10 43.8&15.7&0.9&16533&11&229&12&3\\[0.5ex]
65402361 &15 36 32.47&$-16$ 32 16.3&15.7&0.6&10505&31&232&8 &2\\[0.5ex]
72502210 &15 13 49.98&$-08$ 49 51.2&15.7&0.7&5600 &13&16&16 &3\\[0.5ex]
87001111 &15 41 10.37&$-01$ 35 46.1&15.8&0.6&16897&20&223&6 &3\\[0.5ex]
72600488 &15 44 54.15&$-11$ 43 03.4&15.8&0.7&24012&51&397&62&3\\[0.5ex]
87100850 &16 04 28.03&$-01$ 41 33.1&15.9&0.7&10839&10&132&22&3\\[0.5ex]
79702517 &15 11 38.51&$-04$ 14 50.8&15.9&0.8&10978&23&413&11&2\\[0.5ex]
79800965 &15 32 46.84&$-03$ 19 02.3&15.9&0.5&18012&19&303&8 &3\\[0.5ex]
65502359 &15 52 07.79&$-15$ 13 22.9&15.9&0.6&17513&44& -- & -- &4\\[0.5ex]
87002295 &15 30 27.12&+01 08 12.7&16.1&0.6&15068 &30 & -- & -- &4\\[0.5ex]
86902718 &15 10 28.74&+02 12 43.1&16.1&0.5&11391&32&177 &22 &3\\[0.5ex]
72801203 &16 17 51.62&$-10$ 20 59.9&14.6&1.1&8472&29&215 &4 &2\\[0.5ex]
72801189 &16 17 57.70&$-09$ 46 17.2&14.9&0.9&8006&24&209 &2 &2\\
& & & & & \\ [-1.5ex]
\hline
\ & \multicolumn{9}{c}{~}   \\ 
   \end{tabular} \vspace{-.1in}
\end{center}
\end{table*}
\setcounter{table}{4}
\begin{table*}
\begin{center}
\caption {Spectroscopic parameters for the sample galaxies (continued).}
\vspace{.1in}
\begin{tabular}{lrrrrrrrrc} \hline 
\ & \multicolumn{9}{c}{~}   \\ [-1.5ex]
Galaxy & Right Ascension & Declination & $b_J$ & $d$ &
$v_{hel}$ & err$_{v}$ & $\sigma$ & err$_\sigma$ & $Q_{spec}$\\
\ & \multicolumn{9}{c}{~}   \\ [-1.5ex]
\hline   
& & & \\ [-1.5ex]
{\bf Region~2} &&&&&&&&& \\[0.75ex]
NGC 7010 &21 01 55.45&$-12$ 32 17.1&14.3&1.9&8480&19&228 &3 &1\\[0.5ex]
81405781 &20 53 56.12&$-04$ 01 20.4&14.7&1.1&6025&10&156&10 &1\\[0.5ex]
81303405 &20 35 10.59&$-06$ 16 00.6&15.1&1.0&3844 &5 &89 &5 &1\\[0.5ex]
81502978 &21 17 38.08&$-05$ 05 20.4&15.1&0.9&9102 &17&367&9 &3\\[0.5ex]
67200327 &21 44 53.58&$-13$ 24 53.2&15.1&0.8&5418 &13&167&14&1\\[0.5ex]
81303115 &20 36 22.29&$-05$ 08 59.3&15.2&1.2&3797 &10&131&19&1\\[0.5ex]
74402521 &21 48 16.53&$-09$ 30 02.4&15.2&0.9&5380 &20&213&15&1\\[0.5ex]
67003337 &20 52 46.63&$-13$ 39 34.8&15.3&0.9&24515&38&555&7 &3\\[0.5ex]
66901323 &20 39 01.84&$-14$ 01 30.9&15.4&0.8&8409 &18&265&2 &1\\[0.5ex]
81603636 &21 33 27.73&$-03$ 33 44.1&15.4&0.8&8027 &14&257&11&3\\[0.5ex]
74402642 &21 46 50.74&$-10$ 05 44.3&15.5&1.1&23869&10&598&14&2\\[0.5ex]
81405745 &20 53 59.57&$-03$ 32 27.4&15.5&0.6&5979 &11&126&14&1\\[0.5ex]
67001814 &21 00 46.20&$-15$ 10 30.3&15.5&1.1&8455 &16&241&13&1\\[0.5ex]
81503740 &21 14 12.52&$-04$ 42 24.4&15.6&0.8&13765&11&266&22&1\\[0.5ex]
67003179 &20 53 39.80&$-14$ 35 43.1&15.6&0.8&8468 &6&139 &3 &2\\[0.5ex]
81601185 &21 45 01.36&$-06$ 35 43.1&15.6&0.9&6103 &15&153&13&1\\[0.5ex]
81602395 &21 39 15.52&$-06$ 56 13.7&15.6&0.8&6906 &20&158&6 &2\\[0.5ex]
81404709 &20 56 46.08&$-04$ 38 08.9&15.7&0.8&4680 &49& -- &-- & 4\\[0.5ex]
67003696 &20 50 48.81&$-14$ 41 58.9&15.7&0.8&7580 &11&178&6 &1\\[0.5ex]
81602330 &21 39 29.90&$-04$ 26 37.3&15.7&0.7&8225 &25&114&8 &1\\[0.5ex]
67102806 &21 27 46.38&$-16$ 20 28.2&15.7&0.7&9353 &7 &215&4 &1\\[0.5ex]
74400983 &21 41 10.20&$-08$ 18 11.9&15.7&0.9&15305&22&351&4 &1\\[0.5ex]
81404583 &20 57 05.76&$-02$ 50 34.2&15.8&0.8&17209&34&510&16&2\\[0.5ex]
74401483 &21 35 55.30&$-08$ 07 58.1&15.8&0.9&15751&7&348& 3 &2\\[0.5ex]
67201749 &21 37 30.96&$-14$ 27 30.1&15.8&0.6&18358&25&514&13&1\\[0.5ex]
67202822 &21 38 18.48&$-16$ 55 16.6&15.8&0.7&15509&29&469&13&1\\[0.5ex]
67103161 &21 20 15.95&$-16$ 11 58.7&15.8&0.7&13294&19&231&4 &1\\[0.5ex]
67202363 &21 47 30.24&$-17$ 24 35.4&15.8&0.6&10185&10&174&14&1\\[0.5ex]
74300882 &21 15 30.52&$-07$ 32 02.2&15.8&0.7&8270 &10&122&18&2\\[0.5ex]
67001913 &21 00 17.18&$-16$ 59 57.9&15.9&0.6&8716 &17&246&1 &2\\[0.5ex]
81603664 &21 33 17.98&$-03$ 22 20.5&15.9&0.6&2873 &4 &104&3 &1\\[0.5ex]
81301006 &20 46 07.02&$-07$ 10 50.8&15.9&0.7&12441&17&162&15&1\\[0.5ex]
67001942 &21 00 05.85&$-16$ 00 34.5&15.9&0.7&8501 &27&341&7 &1\\[0.5ex]
81602522 &21 38 30.84&$-04$ 31 25.2&15.9&0.8&16219&11&268&14&1\\[0.5ex]
81302133 &20 40 42.55&$-04$ 40 54.1&16.0&0.7&8611&19&140&13 &3\\[0.5ex]
74301907 &21 19 33.98&$-09$ 38 45.1&16.0&0.7&12340&14&297&11&1\\[0.5ex]
74200901 &21 05 07.05&$-10$ 36 38.4&16.0&0.7&8780&19&213 &5 &1\\[0.5ex]
74405800 &21 34 26.16&$-11$ 27 24.7&16.0&0.7&17794&21&266&11&1\\[0.5ex]
81601843 &21 41 57.89&$-06$ 23 11.7&16.0&0.8&15768&31&249&1 &3\\[0.5ex]
81601806 &21 42 05.59&$-07$ 17 17.9&16.0&0.7&5998 &9&121 &15&2\\[0.5ex]
67002128 &20 59 13.51&$-13$ 22 13.3&16.0&0.7&8492 &17&197& 9&2\\[0.5ex]
81501871 &21 22 42.66&$-03$ 19 32.6&16.0&0.6&5640 &20&77&15 &2\\[0.5ex]
81601380 &21 44 05.41&$-05$ 27 38.9&16.0&0.7&6584 &15 &92&11&1\\[0.5ex]
74200951 &21 04 47.84&$-10$ 55 06.0&16.0&0.7&8269 &10&141&12&2\\[0.5ex]
81504194 &21 12 07.77&$-02$ 33 55.1&16.1&0.7&7957 &7 &111&3 &2\\[0.5ex]
67002172 &20 58 59.41&$-13$ 30 36.4&16.1&0.7&8822 &19&254&7 &1\\[0.5ex]
81603573 &21 33 47.15&$-04$ 11 51.4&16.1&0.6&15726&30&195&2 &2\\[0.5ex]
81602191 &21 40 10.46&$-07$ 05 52.5&16.1&0.6&16043&33&321&1 &2\\[0.5ex]
74401426 &21 36 27.77&$-07$ 31 39.3&16.1&0.7&10425&21&49&24 &2\\[0.5ex]
81602252 &21 39 51.84&$-06$ 39 24.5&16.2&0.6&10348&30&134&6 &2\\
& & & & & \\ [-1.5ex]
\hline
\ & \multicolumn{9}{c}{~}   \\ 
   \end{tabular} \vspace{-.1in}
\end{center}
\end{table*}
\setcounter{table}{4}
\begin{table*}
\begin{center}
\caption {Spectroscopic parameters for the sample galaxies (continued).}
\vspace{.1in}
\begin{tabular}{lrrrrrrrrc} \hline 
\ & \multicolumn{9}{c}{~}   \\ [-1.5ex]
Galaxy & Right Ascension & Declination & $b_J$ & $d$ &
$v_{hel}$ & err$_{v}$ & $\sigma$ & err$_\sigma$ & $Q_{spec}$\\
\ & \multicolumn{9}{c}{~}   \\ [-1.5ex]
\hline   
& & & \\ [-1.5ex]
81600601 &21 47 54.17&$-05$ 36 33.7&16.2&0.6&16715&17&325&19&2\\[0.5ex]
67003806 &20 50 17.86&$-16$ 31 30.7&16.2&0.6&12891&21&330&5 &1\\[0.5ex]
81602134 &21 40 27.29&$-07$ 17 52.8&16.2&0.6&15015&31&251&2 &2\\[0.5ex]
74202489 &20 55 31.78&$-07$ 57 12.4&16.2&0.8&23717&30&349&6 &3\\[0.5ex]
67201815 &21 36 19.16&$-14$ 35 25.1&16.3&0.6&17930&23&446&16&2\\[0.5ex]
67201810 &21 36 28.63&$-14$ 32 41.1&16.3&0.5&17972&16&286&2 &2\\[0.5ex]
81600354 &21 49 07.77&$-03$ 13 57.5&16.4&0.6&16461&17&228&13&1\\[0.5ex]
74404851 &21 45 19.53&$-12$ 08 59.7&16.4&0.7&14761&22&175&7&1\\[0.5ex]
81304119 &20 31 56.30&$-06$ 40 17.5&16.4&0.6&24394&37&360&11&1\\[0.5ex]
67001459 &21 02 45.64&$-12$ 34 59.3&16.4&0.5&7940&16&128&10 &1\\[0.5ex]
67001377 &21 03 14.05&$-16$ 00 59.9&16.4&0.6&8511&19&134& 8 &1\\[0.5ex]
67002239 &20 58 37.77&$-13$ 17 52.6&16.4&0.6&9258&12&167&49 &1\\[0.5ex]
74302893 &21 21 43.60&$-11$ 44 05.5&16.4&0.6&9011&20&133&12 &1\\[0.5ex]
81405246 &20 55 21.94&$-04$ 41 41.1&16.4&0.5&24804&50&452&18&1\\[0.5ex]
74402668 &21 46 34.70&$-09$ 51 06.0&16.4&0.8&10281&37&315&11&1\\[0.5ex]
74402837 &21 44 20.85&$-10$ 44 57.8&16.4&0.5&10820&14&193&8 &1\\[0.5ex]
81400846 &21 08 17.34&$-03$ 58 52.3&16.4&0.5&9427&5&317 & 9 &1\\[0.5ex]
74400638 &21 44 34.04&$-08$ 37 23.9&16.4&0.6&17950&22&259&17&1\\[0.5ex]
67102743 &21 28 57.59&$-16$ 18 04.8&16.4&0.5&18655&17&320&15&1\\[0.5ex]
67201875 &21 34 48.75&$-14$ 55 07.4&16.4&0.5&11424&17&297&34&1\\[0.5ex]
81406413 &20 52 10.94&$-03$ 31 43.4&16.4&0.5&5823&17&51 &15 &1\\[0.5ex]
66900782 &20 44 16.43&$-14$ 09 03.1&16.4&0.6&8742&4&110 & 9 &1\\[0.5ex]
67100128 &21 29 20.05&$-13$ 59 28.8&16.4&0.6&17408&13&322&17&1\\[0.5ex]
88920974 &21 53 26.20&$-01$ 45 11.9& -- &1.0&8045&17&323 &14&1\\[0.5ex]
88520340 &20 44 50.30&+00 06 58.1& -- &2.0&3804 &16&172& 15 &1\\[0.5ex]
88720594 &21 11 23.79&+02 21 24.1& -- &1.9&14544&37&154 &10 &3\\[0.5ex]
88520337 &20 44 36.89&+00 10 41.1& -- &0.7&3865 &7 &41 & 31 &3\\[0.5ex]
{\bf Region~3} &&&&&&&&& \\[0.75ex]
68100721 &00 41 02.46&$-08$ 27 37.1&14.4&1.2&5960&20&142&14 &1\\[0.5ex]
60800316 &00 28 18.95&$-12$ 59 46.4&15.0&0.9&6137&21&73 &24 &1\\[0.5ex]
68101448 &00 32 08.14&$-11$ 02 30.3&15.0&0.8&6210&18&154&13 &1\\[0.5ex]
68001213 &00 15 02.00&$-09$ 49 15.6&15.1&1.0&6718&23&169&7 &1\\[0.5ex]
68101358 &00 33 27.93&$-10$ 23 49.2&15.1&0.8&5982&14&195&12 &1\\[0.5ex]
68101582 &00 30 00.82&$-12$ 20 25.9&15.1&1.1&16971&5&155&5 &2\\[0.5ex]
61100299 &01 25 02.07&$-13$ 59 59.2&15.2&1.0&9894&18&212&11 &1\\[0.5ex]
68100166 &00 48 45.71&$-08$ 52 07.3&15.2&0.8&4282&15&168&11&1\\[0.5ex]
68301163 &01 13 59.64&$-08$ 14 26.4&15.2&0.7&5552&16 &165&12&1\\[0.5ex]
61001080 &01 06 22.24&$-15$ 40 21.3&15.3&1.0&15911&15&263&21&1\\[0.5ex]
68001229 &00 14 54.03&$-09$ 48 27.0&15.3&1.1&6823 &9 &118&14&1\\[0.5ex]
68100890 &00 39 18.72&$-09$ 34 38.6&15.4&0.8&16690&24&354&10&2\\[0.5ex]
68101426 &00 32 25.81&$-09$ 37 03.4&15.4&0.7&6969&11&148&11 &1\\[0.5ex]
68300617 &01 22 47.56&$-12$ 10 25.5&15.4&0.8&15021&28&257&15&1\\[0.5ex]
61001389 &00 57 46.77&$-15$ 34 06.1&15.4&0.9&16567&26&341&6 &2\\[0.5ex]
68100622 &00 42 27.48&$-09$ 09 47.5&15.4&0.7&5804&3 &203&21 &3\\[0.5ex]
68101215 &00 35 26.55&$-08$ 20 55.3&15.5&0.9&11214&34&204&9 &2\\[0.5ex]
60800542 &00 26 07.72&$-13$ 17 11.2&15.6&0.7&7778&12&223&12 &1\\[0.5ex]
61002238 &00 56 22.55&$-16$ 44 20.3&15.6&0.8&16284&28&339&13&3\\[0.5ex]
68300547 &01 23 55.82&$-08$ 49 25.6&15.6&0.7&5253&13&87 &11 &2\\[0.5ex]
68201452 &00 55 39.28&$-08$ 29 15.7&15.6&1.1&4602 &8&178&26 &1\\[0.5ex]
61001909 &01 06 21.72&$-16$ 12 47.6&15.7&0.9&12597&21&345&8 &2\\[0.5ex]
68101346 &00 33 36.91&$-07$ 58 26.3&15.7&0.6&11191&13&245&15&2\\
& & & & & \\ [-1.5ex]
\hline
\ & \multicolumn{9}{c}{~}   \\ 
   \end{tabular} \vspace{-.1in}
\end{center}
\end{table*}
\setcounter{table}{4}
\begin{table*}
\begin{center}
\caption {Spectroscopic parameters for the sample galaxies (continued).}
\vspace{.1in}
\begin{tabular}{lrrrrrrrrc} \hline 
\ & \multicolumn{9}{c}{~}   \\ [-1.5ex]
Galaxy & Right Ascension & Declination & $b_J$ & $d$ &
$v_{hel}$ & err$_{v}$ & $\sigma$ & err$_\sigma$ & $Q_{spec}$\\
\ & \multicolumn{9}{c}{~}   \\ [-1.5ex]
\hline   
& & & \\ [-1.5ex]
60901605 &00 31 47.85&$-12$ 55 55.0&15.8&0.7&8030&16 &265&13&1\\[0.5ex]
61002004 &01 03 47.71&$-16$ 48 34.6&15.8&0.8&11242&29&276&18&1\\[0.5ex]
61100351 &01 23 38.67&$-13$ 40 53.1&15.8&0.9&14201&26&326&3 &2\\[0.5ex]
68201703 &00 53 20.82&$-10$ 15 22.6&15.8&0.8&16356&9&209&12 &2\\[0.5ex]
60800452 &00 27 00.10&$-14$ 16 10.3&15.9&0.7&16200&14&203&14&2\\[0.5ex]
68201935 &00 51 05.05&$-08$ 31 35.7&15.9&0.8&13297&28&275&4 &1\\[0.5ex]
68101515 &00 31 01.35&$-09$ 28 13.5&15.9&0.6&5543&17 &31&23 &2\\[0.5ex]
68100388 &00 45 20.77&$-12$ 03 50.9&15.9&0.5&5886&14&87 & 6 &1\\[0.5ex]
68101166 &00 36 02.25&$-09$ 23 15.9&16.0&0.5&6094&20&44 &21 &1\\[0.5ex]
68201775 &00 52 43.24&$-11$ 31 48.3&16.0&0.7&16375&20&273&13&1\\[0.5ex]
68101270 &00 34 38.49&$-09$ 43 55.1&16.0&0.6&15877&18&141&18&3\\[0.5ex]
68100879 &00 39 21.80&$-11$ 06 24.8&16.0&0.6&15860&19&199&3 &1\\[0.5ex]
61102514 &01 16 52.44&$-16$ 54 12.4&16.1&0.6&6061&8&110 &13 &2\\[0.5ex]
68201336 &00 57 07.98&$-09$ 09 03.2&16.1&0.7&16614&23&213&15&1\\[0.5ex]
61000670 &00 53 34.10&$-13$ 37 41.5&16.1&0.6&16804&26&408&6 &1\\[0.5ex]
61100750 &01 14 38.10&$-14$ 15 39.3&16.1&0.6&15411&29&304&13&1\\[0.5ex]
68201473 &00 55 25.46&$-09$ 47 42.8&16.1&0.6&15060&23&210&17&1\\[0.5ex]
68101018 &00 37 47.50&$-07$ 43 43.7&16.1&0.6&3692&5 &6  & 6 &1\\[0.5ex]
61100580 &01 18 30.28&$-14$ 06 42.2&16.2&0.6&15384&7&349&18 &1\\[0.5ex]
68201394 &00 56 21.84&$-09$ 02 42.6&16.3&0.5&8055 &7&133&13 &1\\[0.5ex]
68100760 &00 40 38.46&$-10$ 08 07.2&16.3&0.5&15142&7&220 &5 &3\\[0.5ex]
68300512 &01 24 25.53&$-07$ 37 57.6&16.3&0.5&10191&32&344&3 &1\\[0.5ex]
61100364 &01 23 19.43&$-12$ 46 09.6&16.4&0.5&10262&24&218&3 &1\\[0.5ex]
61101581 &01 18 08.13&$-14$ 07 33.0& -- & -- &14708&3&246& 12&1\\[0.5ex]
61101582 &01 18 06.14&$-14$ 07 33.6& -- & -- &16458&10&236&18&1\\[0.5ex]
61101583 &01 17 52.50&$-14$ 09 06.4& -- & -- &15292&19&224&8 &1\\[0.5ex]
61101584 &01 17 52.46&$-14$ 14 09.7& -- & -- &20954&5&326 &11&1\\[0.5ex]
61001081 &01 06 29.38&$-15$ 41 12.2& -- & -- &17451&28&454&10&2\\[0.5ex]
61001082 &01 06 21.22&$-15$ 41 17.3& -- & -- &16530&29&170&13&2\\[0.5ex]
61001084 &01 06 25.64&$-15$ 42 15.1& -- & -- &16913&12&228&8 &2\\[0.5ex]
61001085 &01 06 33.34&$-15$ 46 50.8& -- & -- &16300 &10 &158 &17 &3\\
& & & & & \\ [-1.5ex]
\hline
\ & \multicolumn{9}{c}{~}   \\ 
   \end{tabular} \vspace{-.1in}
\end{center}
\end{table*}
% table 6
\begin{table*}
\begin{center}
\caption {Spectroscopic parameters for the standard galaxies.}
\vspace{.1in}
\begin{tabular}{lrrrrrrc} \hline 
\ & \multicolumn{7}{c}{~}   \\ [-1.5ex]
Galaxy & Right Ascension & Declination & $v_{hel}$ & err$_{v}$ & $\sigma$ &
err$_\sigma$ & $Q_{spec}$\\
\ & \multicolumn{7}{c}{~}   \\ [-1.5ex]
\hline   
& & & \\ [-1.5ex]
NGC 5831 &15 01 34&+01 24 54& 1648 & 19 & 139 & 3 & 1\\[0.5ex]
NGC 5845 &15 03 28&+01 49 36& 1383 & 15 & 196 & 2 & 1\\[0.5ex]
NGC 5846A&15 03 56&+01 47 12& 1688 & 18 & 232 & 3 & 1\\[0.5ex]
NGC 6020 &15 55 00&+22 33 00& 4307 & 16 & 173 &15 & 1\\[0.5ex]
NGC 6051 &16 02 48&+24 03 54& 9578 & 17 & 347 &13 & 2\\[0.5ex]
NGC 6086 &16 10 36&+29 36 54& 9549 & 25 & 313 &15 & 1\\[0.5ex]
NGC 7619 &23 17 43&+07 56 00& 3753 & 14 & 269 &14 & 1\\[0.5ex]
NGC 7626 &23 18 10&+07 56 36& 3431 & 15 & 313 & 8 & 1\\[0.5ex]
NGC 227  &00 43 03&$-01$ 48 18&5434 &16 & 243 &15 & 1\\[0.5ex]
IC 1696  &01 22 19&$-01$ 52 42&5860 &11 & 166 & 8 & 2\\[0.5ex]
NGC 533  &01 22 57&+01 30 00& 5521 & 13 & 224 &20 & 1\\[0.5ex]
NGC 636  &01 36 36&$-07$ 45 54&1861 & 19 & 155 & 3& 1\\[0.5ex]
NGC 7468 &23 00 30&+16 20 00&2094 & 22 & 137 & 40 & 2\\[0.5ex]
NGC 7751 &23 44 24&+06 35 00& 3247 & 10 & 123 & 3 & 2\\[0.5ex]
NGC 541  &01 23 12&$-01$ 38 00&5401 &20 & 186 & 14& 1\\[0.5ex]
NGC 545  &01 23 24&$-01$ 36 00&5296 &18 & 203 & 14& 2\\[0.5ex]
NGC 1550 &04 17 00&+02 18 00&3714 & 23 & 327 & 15 & 1\\[0.5ex]
NGC 1587 &04 28 06&+00 33 00& 3597 & 22 &  239 & 16 & 1\\
& & & & & \\ [-1.5ex]
\hline
\ & \multicolumn{7}{c}{~}   \\ 
   \end{tabular} \vspace{-.1in}
   \end{center}
\end{table*}
% table 7
\begin{table*}
\begin{center}
\caption {Spectroscopic parameters for the Coma galaxies.}
\vspace{.1in}
\begin{tabular}{lrrrrrrc} \hline 
\ & \multicolumn{7}{c}{~}   \\ [-1.5ex]
Galaxy & Right Ascension & Declination & $v_{hel}$ & err$_{v}$ & $\sigma$ &
err$_\sigma$ & $Q_{spec}$\\
\ & \multicolumn{7}{c}{~}   \\ [-1.5ex]
\hline   
& & & \\ [-1.5ex]
Coma ~31&12 54 59.0&+27 46 00&7369 & 15 &256 &14 & 1 \\[0.5ex]
Coma ~46&12 55 07.5&+27 52 55&6074 & 12 &210 &13 & 1 \\[0.5ex]
Coma ~49&12 59 29.9&+27 53 36&7893 & 15 &267 &11 & 1 \\[0.5ex]
Coma ~69&12 56 43.5&+28 03 16&7072 & 29 &265 &39 & 3 \\[0.5ex]
Coma ~70&12 56 42.8&+28 02 21&6336 &  7 &187 &27 & 2 \\[0.5ex]
Coma ~72&12 56 27.2&+28 03 33&5656 & 14 &199 & 4 & 2 \\[0.5ex]
Coma ~87&12 57 05.9&+28 03 45&7825 &  6 & 66 &37 & 3 \\[0.5ex]
Coma ~88&12 57 04.5&+28 07 09&6790 & 19 &236 &16 & 3 \\[0.5ex]
Coma 103&12 57 06.0&+28 09 13&4744 &  8 &205 & 4 & 1 \\[0.5ex]
Coma 105&12 56 58.7&+28 10 59&5104 & 25 &170 &10 & 2 \\[0.5ex]
Coma 106&12 56 58.1&+28 10 05&5087 &  8 &199 &11 & 2 \\[0.5ex]
Coma 107&12 56 55.6&+28 09 25&6452 &  9 &138 & 1 & 3 \\[0.5ex]
Coma 118&12 58 15.2&+28 11 42&7552 & 28 &141 & 8 & 1 \\[0.5ex]
Coma 124&12 57 19.8&+28 11 02&6678 &  4 &236 & 6 & 1 \\[0.5ex]
Coma 125&12 57 17.9&+28 11 47&6900 &  3 &174 & 5 & 2 \\[0.5ex]
Coma 129&12 57 11.1&+28 13 53&7223 & 11 &266 &11 & 3 \\[0.5ex]
Coma 130&12 57 09.6&+28 13 09&7209 &  8 &234 & 1 & 2 \\[0.5ex]
Coma 133&12 56 50.6&+28 14 33&4859 &  7 &257 & 5 & 1 \\[0.5ex]
Coma 134&12 56 39.4&+28 13 51&6978 &  6 &184 & 1 & 2 \\[0.5ex]
Coma 135&12 56 35.7&+28 14 23&8286 & 33 &137 &14 & 2 \\[0.5ex]
Coma 136&12 56 30.6&+28 14 12&5682 & 14 &266 & 9 & 2 \\[0.5ex]
Coma 143&12 58 30.1&+28 16 42&4990 & 15 &217 &13 & 1 \\[0.5ex]
Coma 148&12 57 43.7&+28 14 54&6360 & 22 &431 &20 & 1 \\[0.5ex]
Coma 150&12 57 42.0&+28 16 33&7241 &  5 &202 &15 & 3 \\[0.5ex]
Coma 151&12 57 40.0&+28 15 32&6350 &  5 &157 &21 & 2 \\[0.5ex]
Coma 153&12 57 19.2&+28 15 57&6912 & 90 &215 & 9 & 1 \\[0.5ex]
Coma 159&12 56 48.4&+28 14 54&6881 & 20 &252 & 4 & 1 \\[0.5ex]
Coma 160&12 56 41.0&+28 15 58&7672 & 14 &153 &13 & 1 \\[0.5ex]
Coma 168&12 58 24.5&+28 21 40&7088 & 18 &226 &23 & 3 \\[0.5ex]
Coma 172&12 57 50.3&+28 18 45&5827 &  9 &211 &15 & 3 \\[0.5ex]
Coma 179&12 56 53.1&+28 21 13&4638 & 16 &234 &15 & 1 \\[0.5ex]
Coma 180&12 56 49.3&+28 20 53&7780 & 11 &104 & 6 & 3 \\[0.5ex]
Coma 193&12 57 30.8&+28 24 00&7501 & 20 &126 &12 & 2 \\[0.5ex]
Coma 194&12 56 39.3&+28 23 47&7980 &  2 &285 & 2 & 2 \\[0.5ex]
Coma 206&12 57 50.9&+28 28 17&8507 & 17 &268 &15 & 1 \\[0.5ex]
Coma 207&12 57 43.8&+28 26 31&6779 & 16 & 82 &12 & 2 \\[0.5ex]
Coma 217&12 57 32.8&+28 31 08&6803 &  9 &194 & 4 & 3 \\[0.5ex]
Coma 232&12 58 03.8&+28 36 55&6012 & 11 &163 &13 & 1 \\[0.5ex]
Coma 239&12 55 07.0&+28 44 28&6322 &  3 &205 & 1 & 1 \\[0.5ex]
Coma 240&12 55 09.0&+28 45 06&6834 & 15 &246 & 2 & 1 \\
& & & & & \\ [-1.5ex]
\hline
\ & \multicolumn{7}{c}{~}   \\ 
   \end{tabular} \vspace{-.1in}
   \end{center}
\end{table*}
% figure 9
\begin{figure} 
%\picplace{6cm}
%\centerline{
\psfig{file=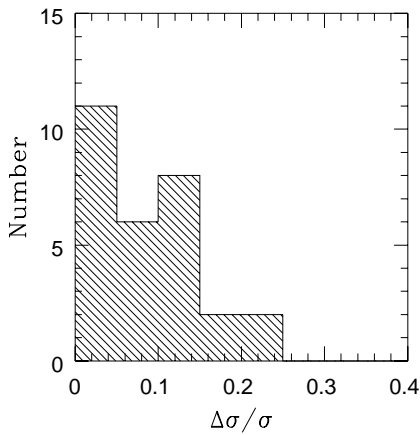,height=60mm}
%}
\caption{Histogram of the fractional errors in the values of $\sigma$
determined from repeat observations.}
\label{delta}
\end{figure}
% figure 10
\begin{figure*} 
%\picplace{5cm}
%\centerline{
\psfig{file=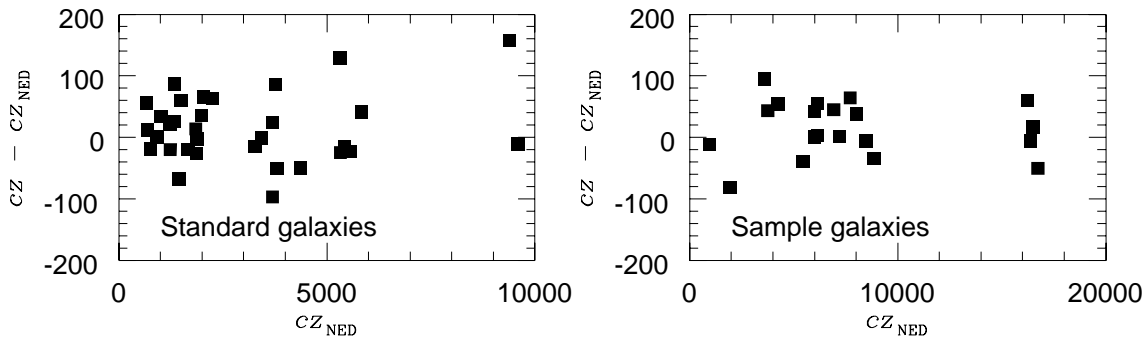,height=50mm}
%}
\caption{Comparison of results for heliocentric radial velocity with published data for 32
standard galaxies and 20 sample galaxies.}
\label{mened}
\end{figure*}
% figure 11
\begin{figure}
%\picplace{8cm}
%\centerline{
\psfig{file=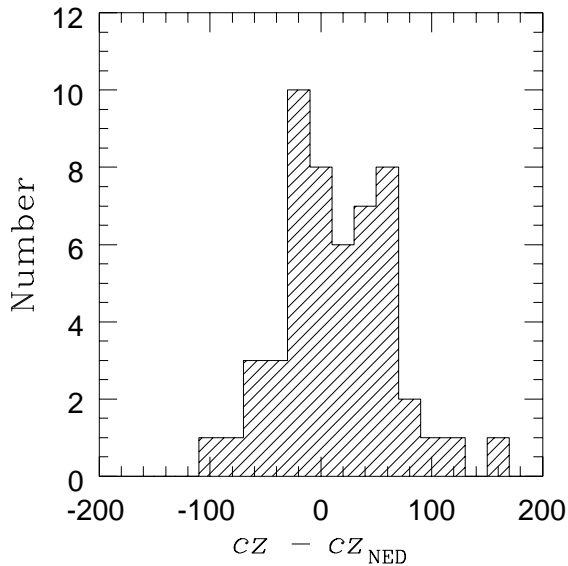,height=80mm}
%}
\caption{ Distribution of differences between measured redshifts and the redshifts given by NED
(literature redshifts), for 52 standard galaxies and sample galaxies.}
\label{distrib}
\end{figure}
% figure 12
\begin{figure*} 
%\picplace{10cm}
%\centerline{
\psfig{file=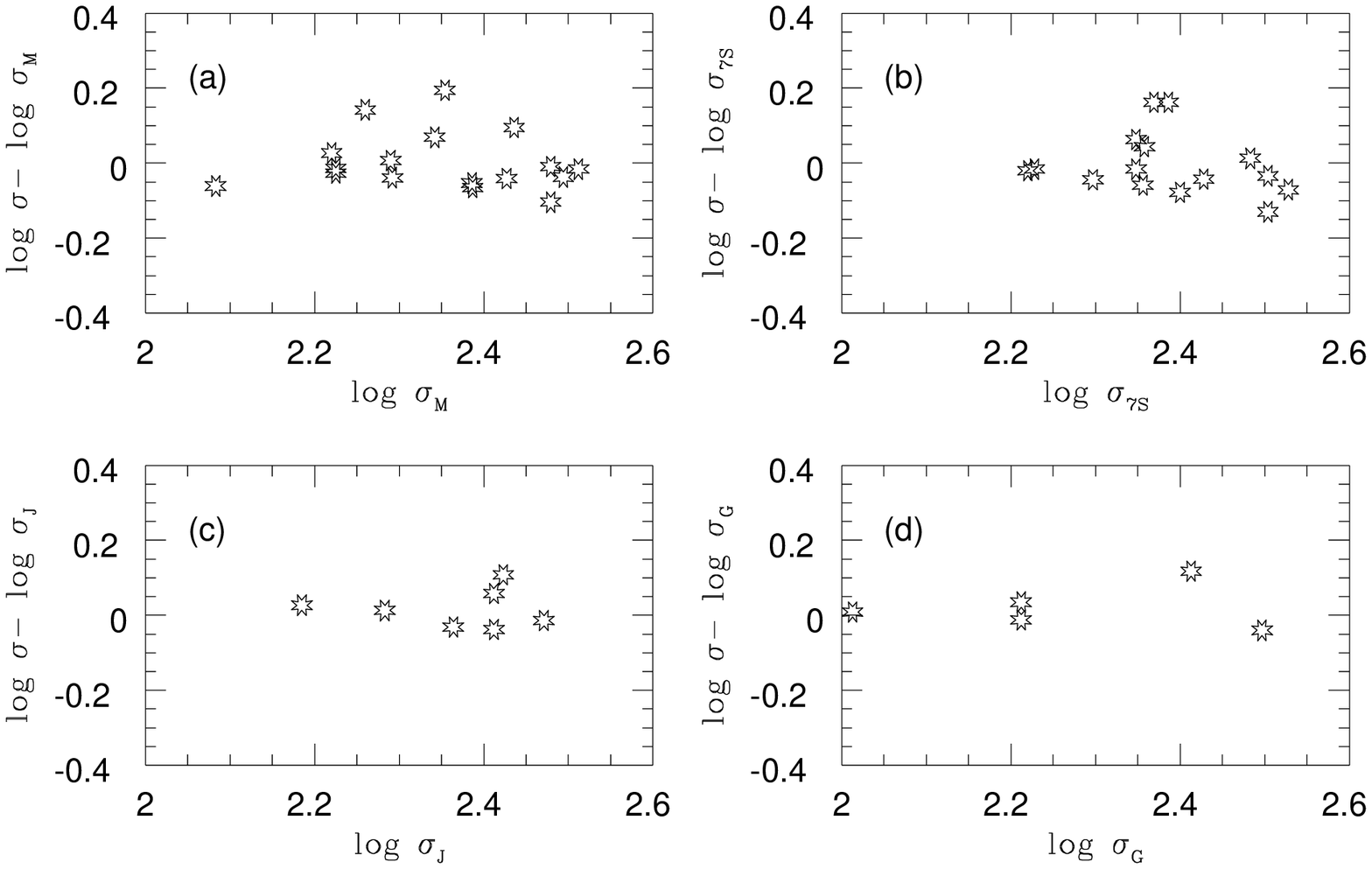,height=100mm}
%}
\caption{ Comparison of the results for velocity dispersions of standard galaxies with values
from the literature: 
(a) McElroy (1995); 
(b) Davies $et$ $al.$ (1987); 
(c) J$\o$rgensen $et$ $al.$ (1995); 
(d) Gonz\'{a}lez (1993).}
\label{external}
\end{figure*}
\subsection{Results for Spectroscopic Parameters}
The results for the spectroscopic parameters are presented in Tables~ 5, 6, and 7. In cases where
there were two or more spectra of the same galaxy available, the results with the smallest errors,
from the spectrum with the highest $S/N$, were used. Histograms of the redshift distributions
in the three sample regions are plotted in Fig.~5. 

Spectroscopic results for the sample galaxies are given in Table~5.
The columns in the table are as follows:\\
1. Galaxy number in the APM catalog (or the galaxy name from the NGC or CGCG catalog in
a few cases); the first three digits of the APM galaxy number are the UK Schmidt plate number.\\
2 \& 3. Right Ascension and Declination (epoch J1950.0) taken from the APM catalog\\
4 \& 5. Magnitude $b_J$ (in mag) and major-axis diameter $d$ (in arcmin) from the APM
catalog (both values corrected for Galactic extinction)\\
6. Heliocentric redshift $v_{hel}$ (in km~s$^{-1}$)\\
7. Estimated internal error in $v_{hel}$ (in km~s$^{-1}$)\\
8. Central velocity dispersion $\sigma$ (in km~s$^{-1}$)\\
9. Estimated internal error in $\sigma$ (in km~s$^{-1}$)\\
10. Quality rating $Q_{spec}$ for the spectrum.

In Table~5, the galaxies are ordered by magnitude, with a few exceptions. The galaxies
CGCG2158 (in Region 1) and NGC7010 (in Region 2) were added for completeness. In Region
1, the galaxy denoted 87101109E is the elliptical galaxy close to 87101109, which was found
to be a spiral. Also included are some galaxies from plates which are not in the final APM
catalog: two galaxies from plate 728 in Region 1, and four galaxies (from plates 885, 887, and
889) in Region 2 (no magnitudes are given for these galaxies, since the magnitude zero-pointing
and plate-matching calibration was not done for these plates). 
In Region 3, there are eight extra galaxies: a group near 61101581 and a group near 61101081,
for which spectra had been obtained, and which were included in the sample.

The results for the spectroscopic parameters of the standard galaxies and Coma galaxies are
presented in Tables 6 and 7 respectively. The columns are as in Table 5, except that magnitudes
and major-axis diameters are not given. The names of the galaxies in Coma are taken from the
Dressler numbers as catalogued by Dressler (1980).
% section 7
\section{Internal and External Comparisons}
\subsection{Internal Comparisons}
The reliability of the determinations of the spectroscopic parameters was checked by comparing
the results from repeat observations of the same galaxy. There are 30 galaxies of which two
spectra were obtained for each, and for these galaxies the differences in parameters were found. 
\subsubsection{Redshifts}
Internal comparisons of radial velocities are shown in Fig.~6 for 30 galaxies with two spectra
each.
% more around 5-8000 km/s because more repeats for Coma
The order ($cz_1$ and $cz_2$) in which the two results have been considered is arbitrary.
The rms scatter is $\Delta cz_{hel}$ = 40.7 km~s$^{-1}$
for the comparisons of repeat observations.
\subsubsection{Velocity dispersions}
In Fig.~7, internal comparisons of velocity dispersions are shown for 29 galaxies with two
spectra each.
%aperture corrections done before comparisons
An estimate of the accuracy is the rms scatter of 
$|\log \sigma_1$ - $\log \sigma_2|$, which is 0.045.
In Fig.~8, the errors in repeat measurements of $\sigma$ are also shown as a function of $S/N$
per $\mbox{\AA}$.
It is clear from the plot that the repeatability depends on the $S/N$ of the spectrum.
For a $S/N$ of 40 -- 50, velocity dispersions can be determined to within 4 -- 5\,\% (from $\Delta
\sigma/\sigma$) and for a $S/N$ of $\sim$ 20 (which is the case for most of the galaxies in the
sample) the accuracy is about 8 -- 10\,\%.

Fig.~9 is a histogram of the fractional errors in $\sigma$ for 29 repeat observations. 
The rms scatter in $\Delta \sigma/\sigma$ is 0.103, 
which means that $\sigma$ can be determined to 10\,\% accuracy.
The galaxies for which there are repeat measurements are typical of those in the whole sample,
so these estimated errors are representative of the true errors. 
\subsection{External Comparisons}
\subsubsection{Redshifts}
Results for redshifts were compared with values 
from NED\footnote{NED, the NASA/IPAC Extragalactic Database, 
is operated by the Jet Propulsion Laboratory, California Institute of Technology under contract
with the National Aeronautics and Space Administration.} (Madore $et$ $al.$ 1992) and from
the ZCAT compilation of redshifts (Huchra $et$ $al.$ 1992), and the differences were found to
be relatively small. Redshift comparisons were possible for 32 standard galaxies and 20 sample
galaxies. (Redshifts were available for only a small fraction of the galaxies in the sample
regions, since the South Equatorial Strip had not been well studied before.)

The determinations of radial velocity for these two sets of galaxies were compared with redshifts
from the literature,
and the results are shown in Fig.~10. For both data sets the agreement is good. 
For all 52 galaxies for which literature redshifts were available, the distribution of differences
between the measured redshifts and the literature values is shown in the histogram of Fig.~11.
The values of the mean absolute difference 
$|cz - cz_{\mbox{NED}}|$ are 42.4 km~s$^{-1}$ and 37.3 km~s$^{-1}$ for the standard galaxy
and sample galaxy comparisons respectively; the rms differences are 56.0 km~s$^{-1}$ and 45.8
km~s$^{-1}$. This shows that the redshifts are of adequate reliability.
\subsubsection{Velocity Dispersions}
In order to make sure that the values of velocity dispersions determined here are scaled to the
standard system, the results for the velocity dispersions of standard galaxies were compared with
published values. Our data set overlapped with the following sets:
(i)~the compilation of McElroy (1995) (17 galaxies in common);
(ii)~Davies $et$ $al.$ (1987) (7S) (15 galaxies in common); 
(iii)~J$\o$rgensen $et$ $al.$ (1995) (7 galaxies in common); and 
(iv)~Gonz\'{a}lez (1993) (5 galaxies in common).

The data set of Davies $et$ $al.$ (the 7S data) defines a good standard system 
since it is frequently used for comparisons.
The comparison with the results of Davies $et$ $al.$ is shown
in Fig.~12 together with comparisons with data from the other sources.
The difference between the values from this study and the literature data,
$\Delta \log \sigma = \log \sigma$(this study)$ - \log \sigma$(literature),  
are shown plotted against $\log \sigma$(literature).
% aperture correction described above was applied to lit data
There do not appear to be any offsets or systematic differences relative to the velocity
dispersions of the four data sets. 
The rms scatter of $\Delta \log \sigma$ is 0.066, 0.111, 0.036, and 0.021 for the plots in (a), (b),
(c), and (d) respectively.
The agreement is best for comparisons with Gonz\'{a}lez (1993) and J$\o$rgensen $et$ $al.$
(1995), which are the most recent data sets, although the numbers of galaxies in common are
smaller.
% section 8
\section{Summary}
We have obtained new radial velocities and central velocity dispersions for 179 E and S0
galaxies in three selected directions in the APM South Equatorial Strip, as well as for 
40 galaxies in the Coma cluster. Observations were made with the 2.4~m and 1.3~m telescopes
of the MDM Observatory on Kitt Peak, Arizona, using the Mark III spectrograph, and at the
4.4~m MMT, using the Red Channel. The spectra have a mean $S/N$ per $\mbox{\AA}$ of 23.

Radial velocities and central velocity dispersions have been determined by the Fourier
cross-correlation method. The velocity dispersions have been corrected for the effect of the
aperture size and for the galaxy's effective radius.  We find that the typical uncertainties on the
derived parameters are $\pm$ 40 km~s$^{-1}$ in $cz$ and $\pm$ 0.045 in $\log \sigma$.
In comparisons with literature data, no offsets or systematic differences are seen. 
The accuracy of 8 -- 10\,\% in the derived velocity dispersions is high enough for the values to
be used in the application of Fundamental Plane analysis.
The results given here have been used together with photometric data (M\"uller $et$ $al.$ 1999)
to derive Fundamental Plane distances to the sample galaxies in order to determine their peculiar
velocities and thereby investigate the reality of large-scale streaming motion; results have been
reported in M\"uller $et$ $al.$ (1998).
\begin{acknowledgements}
GW wishes to acknowledge partial support from the Alexander von Humboldt Foundation during
a visit to the Ruhr-Universit\"at Bochum, and also from NSF grant AST93-47714. 
KM acknowledges financial support from a Dartmouth Fellowship and from a 
Research Studentship at the European Southern Observatory in Munich. 
\end{acknowledgements}

\end{document}